\documentclass[a4paper, amsfonts, amssymb, amsmath, reprint, showkeys, nofootinbib,twoside]{revtex4-1}

\usepackage[english]{babel}
\usepackage[colorinlistoftodos, color=green!40, prependcaption]{todonotes}
\usepackage{amsthm}
\usepackage{mathtools}
\usepackage{physics}
\usepackage{xcolor}
\usepackage{graphicx}
\usepackage[left=23mm,right=13mm,top=35mm,columnsep=15pt]{geometry} 
\usepackage{adjustbox}
\usepackage{placeins}
\usepackage[T1]{fontenc}
\usepackage{lipsum}
\usepackage{csquotes}
\usepackage[pdftex, pdftitle={Article}, pdfauthor={Author}]{hyperref} 
\begin{document}
\title{Response of undoped cryogenic CsI to low-energy nuclear recoils  }

\author{C.M. Lewis}
\email{marklewis@uchicago.edu}
\author{J.I. Collar}
\affiliation{Enrico Fermi Institute, Kavli Institute for Cosmological Physics, and Department of Physics, University of Chicago, Chicago, Illinois 60637, USA}

\date{\today}

\begin{abstract}
The bright scintillation of pure CsI operated at liquid-nitrogen temperature makes of this material a promising dark matter and neutrino detector. We present the first measurement of its quenching factor for nuclear recoils. Our findings indicate  it is indistinguishable from that for sodium-doped CsI at room temperature. Additional properties such as light yield, afterglow, scintillation decay properties for electron and nuclear recoils, and energy proportionality are studied over the \mbox{108-165 K} temperature range, confirming the vast  potential of this medium for rare-event searches.  
\end{abstract}

\maketitle

\section{Introduction} \label{sec:introduction}

An ongoing interest in characterizing the response of radiation detectors to low-energy nuclear recoils, induced by the elastic scattering of neutral particles, is traceable to the first direct search for Weakly Interacting Massive Particles (WIMPs) \cite{wimp1}, popular dark matter candidates. Neutrinos with energies below a few tens of MeV can scatter coherently from nuclei via the weak neutral current \cite{freedman}, also producing few-keV nuclear recoils as the single outcome from this process. The recent observation of this so-called Coherent Elastic Neutrino-Nucleus Scattering (CE$\nu$NS) \cite{sciencepaper, Scholz_2018} has added thrust to a quest for new  materials well-adapted to the detection of these subtle low-energy interactions. 

Scintillating sodium-doped cesium iodide (CsI[Na]), operated at room-temperature, was chosen as the favored detector material for the first CE$\nu$NS measurement. A long list of virtues leading to its selection is described in \cite{setupSNS,nicolethesis}. Among those is a large and essentially identical CE$\nu$NS cross-section for both Cs and I, a high light-yield, reduced afterglow, and a quenching factor (QF) of order 10 \% in the few-keV nuclear recoil (NR) energy region of interest. This QF is the ratio between the light yield for NRs and that for electron recoils (ERs) of the same energy. A precise understanding of the energy dependence of the QF is of crucial importance in the interpretation of WIMP and CE$\nu$NS searches \cite{csiqf}. 

Undoped CsI exhibits a large increase in light yield at liquid-nitrogen temperature, reaching a theoretical limit in light-conversion efficiency that exceeds 100 scintillation photons per keV of ER energy deposition \cite{amsler,mos1,mos2,nadeau,clark,liu,woody,zhang,mik}. This is close to three times the room-temperature yield of CsI[Na]. When monitored with silicon light sensors combined with state-of-the-art waveshifters able to maximize their quantum efficiency, the potential to detect NRs as low in energy as 1 keV appears to be within reach \cite{ESS}. This is a NR energy regime unprecedented for scintillators; one where the new physics beyond the Standard Model that is reachable via CE$\nu$NS concentrates \cite{ESS}. However, the assumption on which this promise is based is that the QF of this cryogenic material, unknown for NRs until now, is at least as favorable as for doped CsI at room-temperature.

\begin{figure}
    \centering
    \includegraphics[width=.67\linewidth]{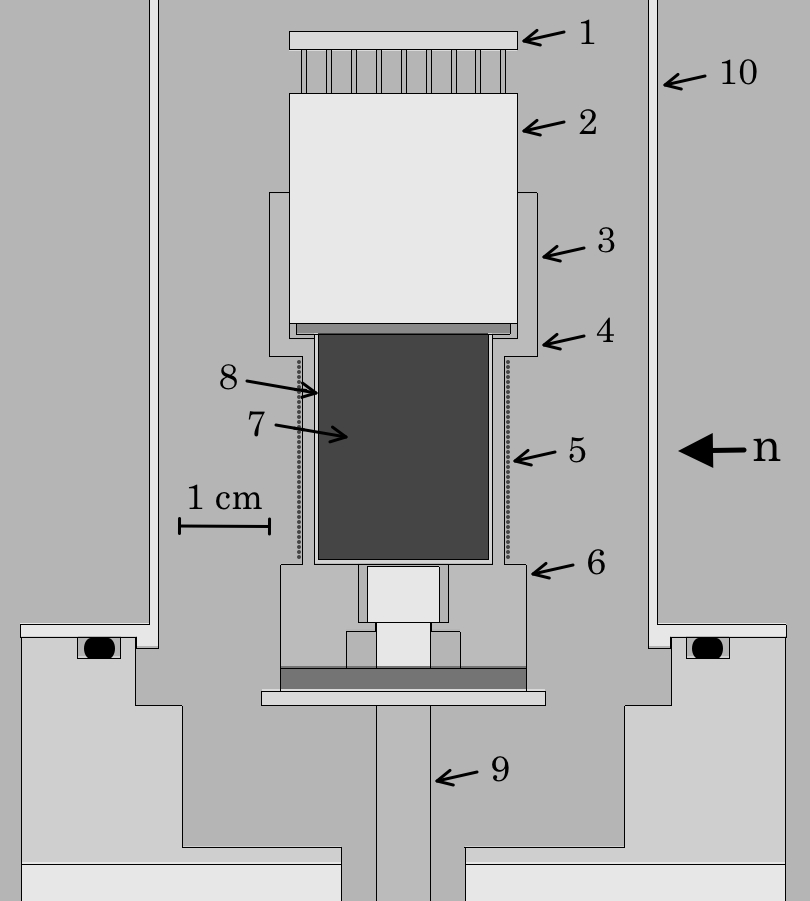}
    \caption{Detector cross-section, derived from the MCNP simulation: 1) voltage divider, 2) Hamamatsu R8520-506 cryogenic PMT, 3) copper holder, 4)   position of thermocouple \#1, 5) coiled manganin wire, 6) thermocouple \#2, 7) CsI crystal, 8) PTFE reflector, 9) cold finger, immersed in liquid-nitrogen Dewar, 10) stainless steel endcap.  A vacuum manifold on the endcap houses cable feedthroughs. The direction of  incoming neutrons is indicated. }
    \label{fig:cryostat}
\end{figure}

This work describes a first measurement of this QF in the temperature range 108-165 K, using the custom detector assembly in Fig.\ \ref{fig:cryostat}, exposed to monochromatic 2.25 MeV neutrons from a D-D generator. Neutrons scattering from the CsI crystal are detected by a Bicron 501A liquid scintillator cell with neutron/gamma discrimination capability. This cell is placed at a user-defined angle from the initial neutron trajectory, allowing to select the energy deposited by NRs in CsI. This experimental setup, data acquisition  system (DAQ), and analysis method have been previously employed by us for NaI[Tl] and CsI[Na] room-temperature QF measurements. Details of these technical aspects are provided in \cite{setupNa,setupSNS,csiqf}. In this new implementation, a PID algorithm was used to monitor the temperature at both ends of the CsI crystal and control the power injected into a heating element (manganin wire, Fig.\ \ref{fig:cryostat}) resulting in a temperature stability of $\sim$0.03 K, and a gradient across the crystal of $<$1.5 K, for all present runs. 

In the second section of this paper we describe our QF measurements. The third section includes the determination of a number of additional quantities (light yield, afterglow, scintillation decay properties for ERs and NRs, and energy proportionality) of interest to assess the potential of this new material for rare event searches. Its extraordinary promise is emphasized in our conclusions.

\section{isolation of low energy nuclear recoils and QF measurement} \label{measurement}

The use of a small 7.24 cm$^3$ CsI scintillator \cite{proteus} ensures that single-scatters dominate neutron interactions in the crystal. Multiple scatters make up for just \mbox{17-27\%} of the total, depending on the selected scattering angle, and are accounted for in simulations. A Hamamatsu R8520-506 cryogenic bialkali photomultiplier (PMT) was directly coupled to the sample using optical RTV.  While operation of this PMT down to \mbox{87 K} is possible, the lowest temperature of 108 K achieved in this study was limited by the cooling power of the horizontal-arm cryostat employed. 

\begin{figure}[!htbp]
    \centering
    \includegraphics[width=.8\linewidth]{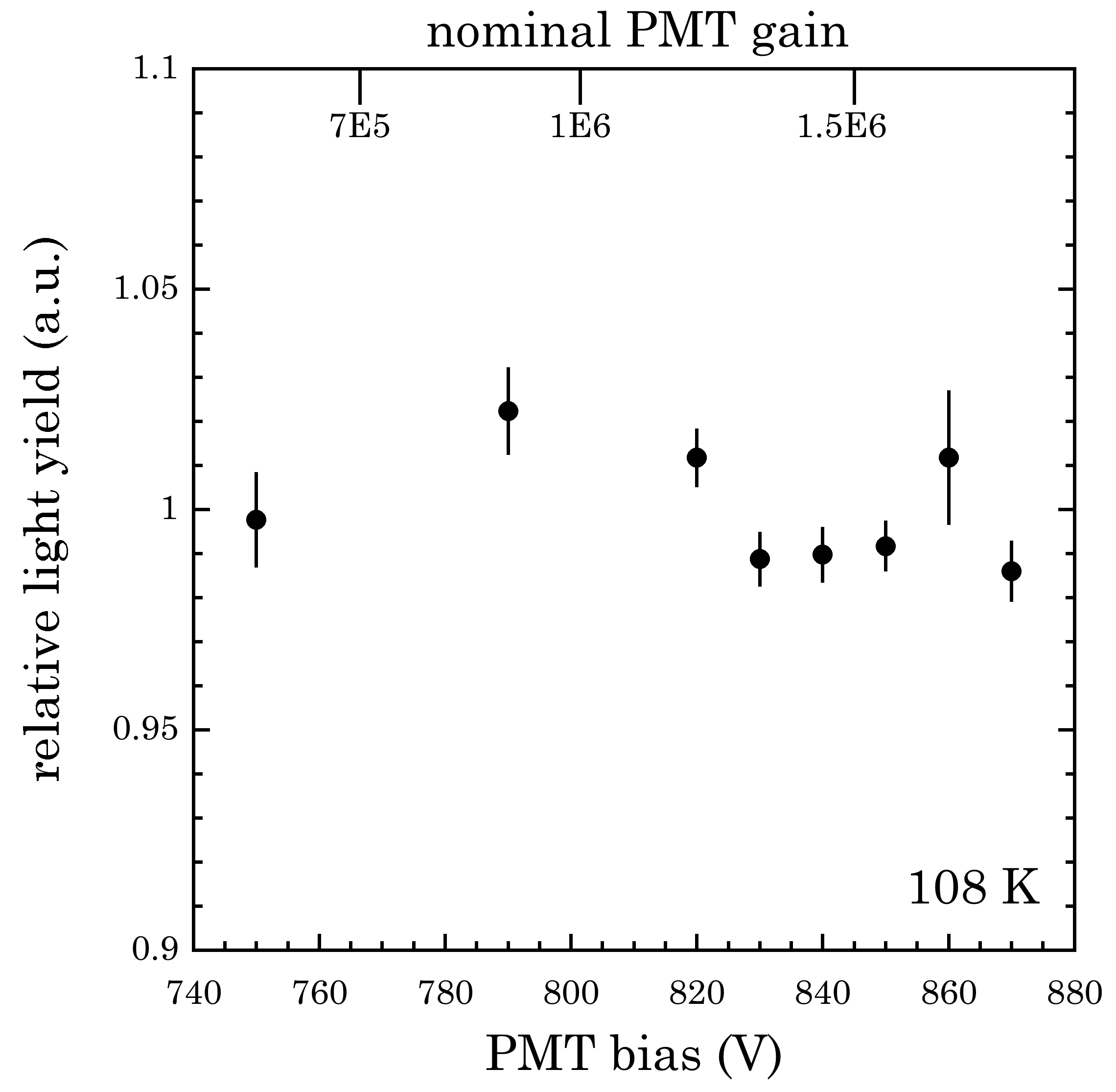}
    \caption{Tests of R8520-506 PMT saturation under 59.5 keV gamma irradiation of the CsI crystal. Light yield is normalized to the average of all measurements. Error bars combine the uncertainties from fits to SPE and 59.5 keV charge distributions \cite{csiqf}. A bias of 820 V was adopted.   }
    \label{fig:saturation}
\end{figure}

Among the lessons derived from our latest CsI[Na] experimentation is the impact on QF measurements of PMT saturation at high bias. As a preliminary precaution, we compared the PMT charge output for $^{241}$Am 59.5 keV gammas and for single photo-electrons (SPE) following the test procedure developed in \cite{csiqf}. The normalized ratio of these outputs provides a light yield in units of PE/keV. As in \cite{csiqf}, the energy reference used in the definition of the QF was given by this $^{241}$Am emission, assuming direct proportionality for lower ER energies. Tests over a range of PMT voltage biases were made at 108 K, a temperature corresponding to the maximum light yield observed in this work. As can be seen in Fig.\ \ref{fig:saturation} the chosen 820 V bias is well within the linear response of this PMT and away from saturation effects at the light levels involved in this study.

\begin{figure}[!htbp]
    \centering
    \includegraphics[width=.8\linewidth]{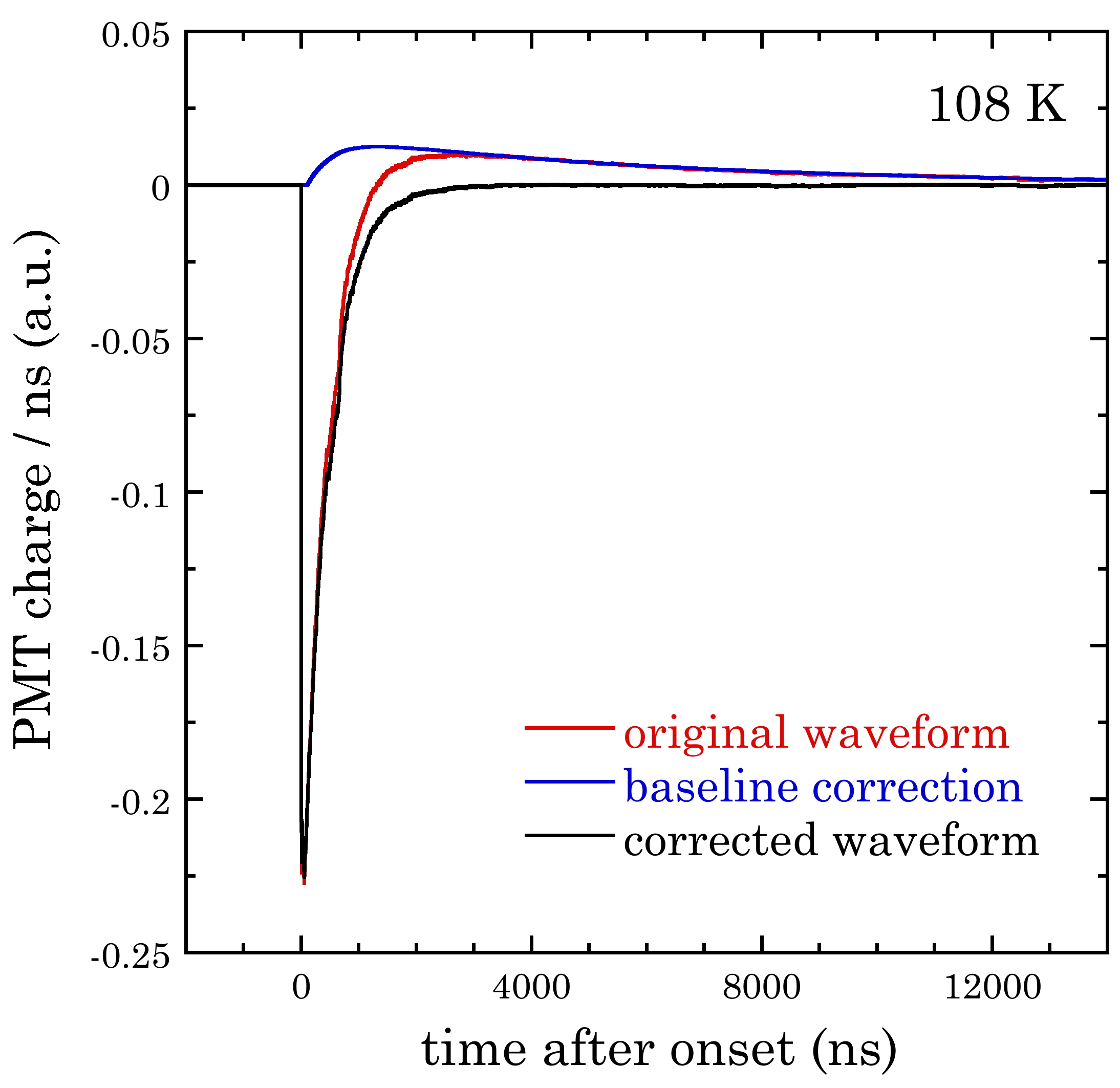}
    \caption{Implementation of an offline baseline correction using an inverse high-pass algorithm on an example co-added ensemble of 1,000 $^{241}$Am gamma signals. The final integrated charge over the first 3 $\mu$s is corrected by $\sim$18$\%$ with respect to the original trace.}
    \label{fig:overshoot}
\end{figure}

\begin{figure}
    \centering
    \includegraphics[width=.8\linewidth]{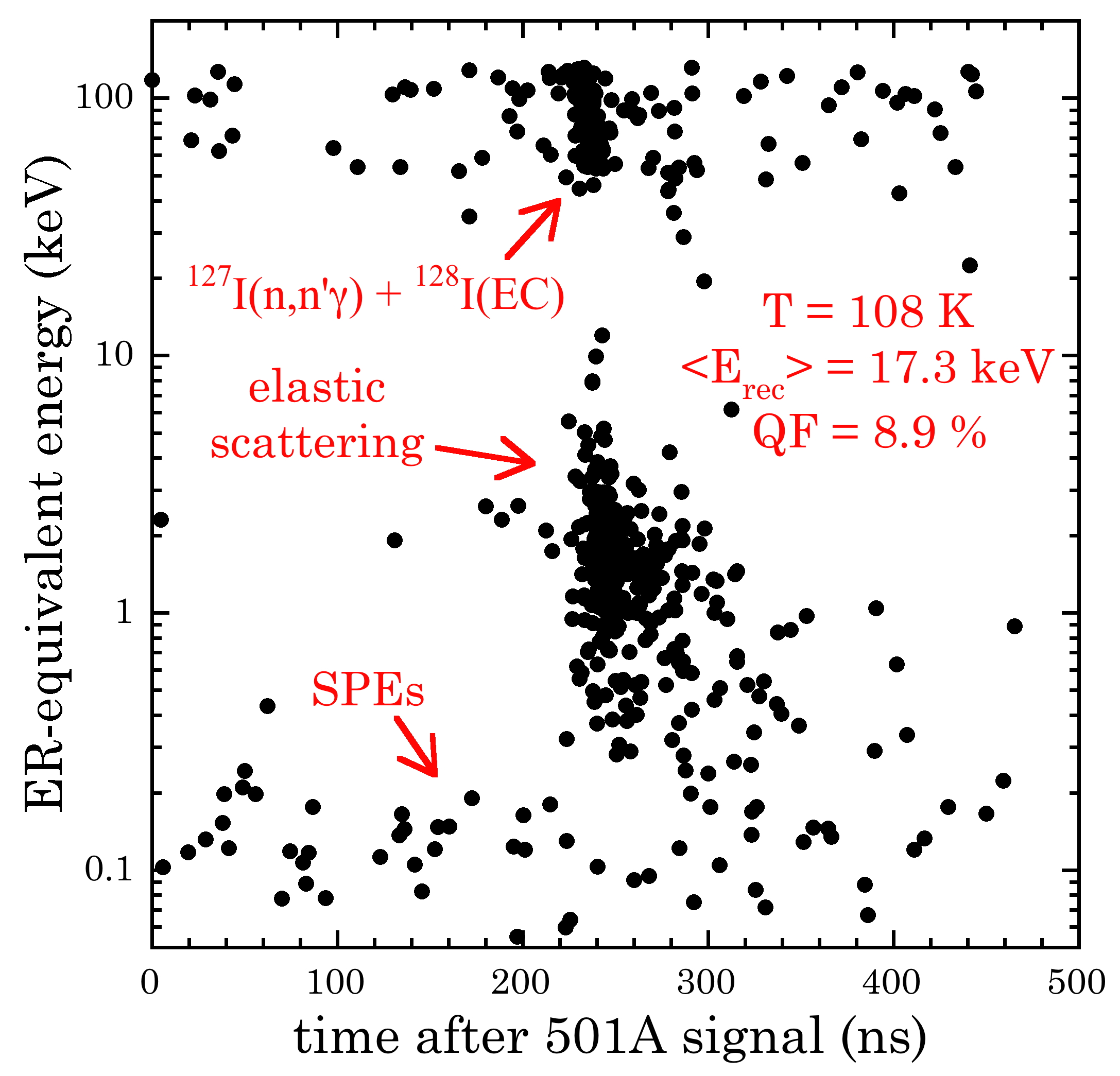}
    \caption{Scatter plot of CsI events passing the Bicron 501A IRT cut for the 65$^\circ$ neutron scattering angle. Prompt coincidences between the backing detector and CsI crystal appear at $\sim$225 ns along the horizontal scale in this DAQ \cite{setupNa}. A scintillation decay time of $\sim$600 ns at 108 K \cite{amsler} results in a modest spillage of the onset of few-PE signals to later times.}
    \label{fig:NScatter}
\end{figure}

Incremental improvements to the MCNP-PoliMi simulations \cite{MCNP-PoliMi} used in \cite{csiqf} were made by accounting for subdominant inelastic neutron scattering through de-exitation gamma escape from CsI. Similarly to \cite{csiqf}, charge was integrated over the 3 $\mu$s following the onset of scintillation signals. However, due to electrical safety concerns for metallic-envelope PMTs like the R8520-506, a positively-biased voltage divider was used. The resulting capacitive coupling to the DAQ produces a well-documented overshoot of the PMT signal \cite{wright2017pmt}. Uncorrected, this leads to an underestimation of the integrated charge carried by a scintillation signal. Remedial analysis techniques have been put forward in a number of affected experiments \cite{boon,chooz,juno}. The impact of this effect and its correction on our charge measurements can be assessed from Fig.\ \ref{fig:overshoot}. The corrective procedure uses an inverse high-pass algorithm (an offline pole-zero cancellation) to allow for accurate charge integration below the median waveform baseline. Special attention was paid to ensure that this average charge correction also applied to signals at lower energy. As expected, since the response of the output capacitor causing the overshoot depends on signal frequency, and not on amplitude, the magnitude of the integrated charge correction was found to be consistent at 18$\%$ for all energies. 

\begin{figure}
    \centering
    \includegraphics[width=.7\linewidth]{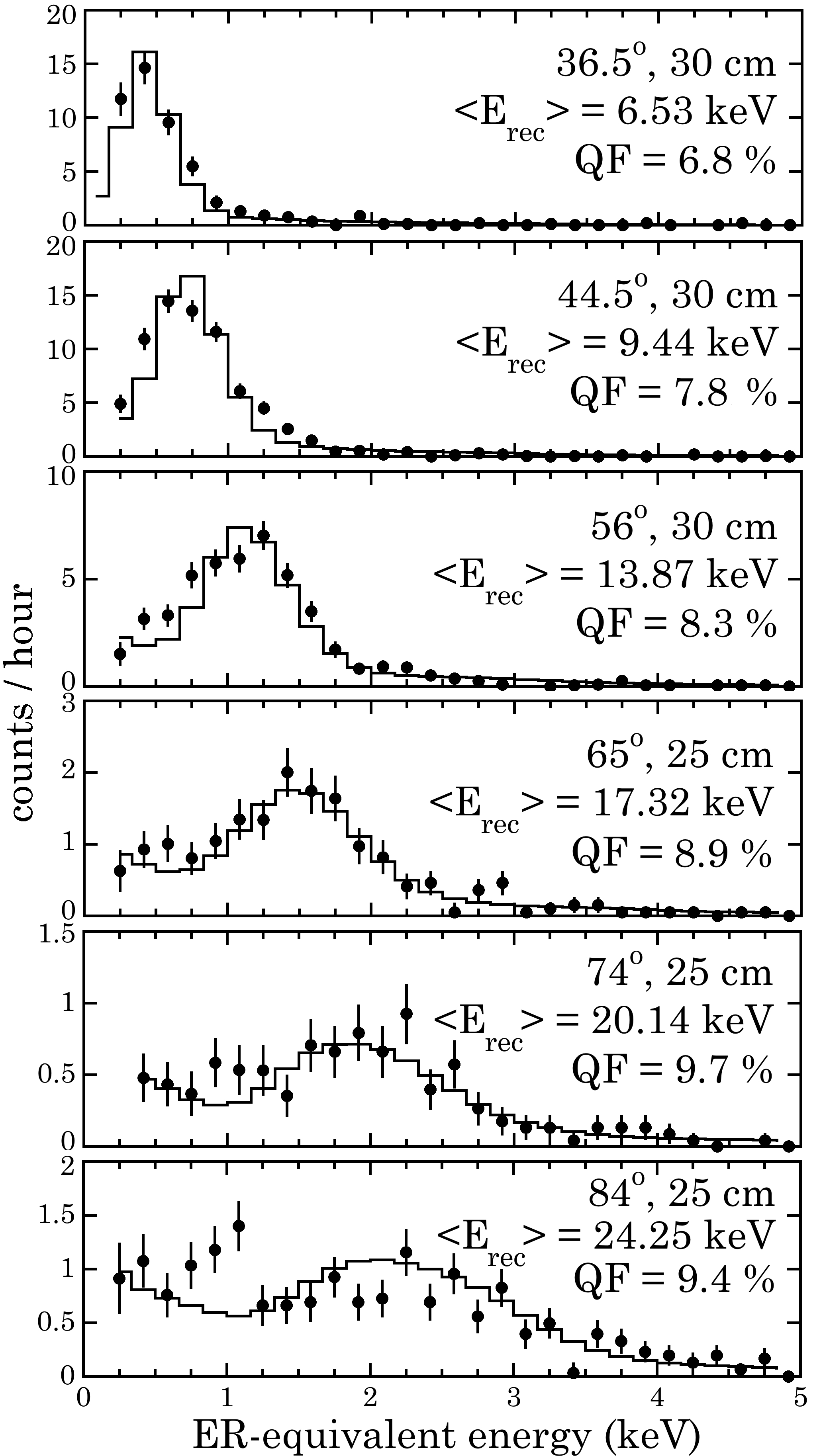}
    \caption{Energy deposition by NRs from neutron scattering on CsI at 108 K. Datapoints are experimental data, histograms correspond to simulated distributions at best-fit QF. Scattering angle, backing detector distance to the CsI crystal, simulated mean NR energy, and best-fit QF are indicated. The decrease in event rate with  increase in angle is characteristic of forward-peaked elastic scattering.}
    \label{fig:108K_layout}
\end{figure}

\begin{figure}
    \centering
    \includegraphics[width=.8\linewidth]{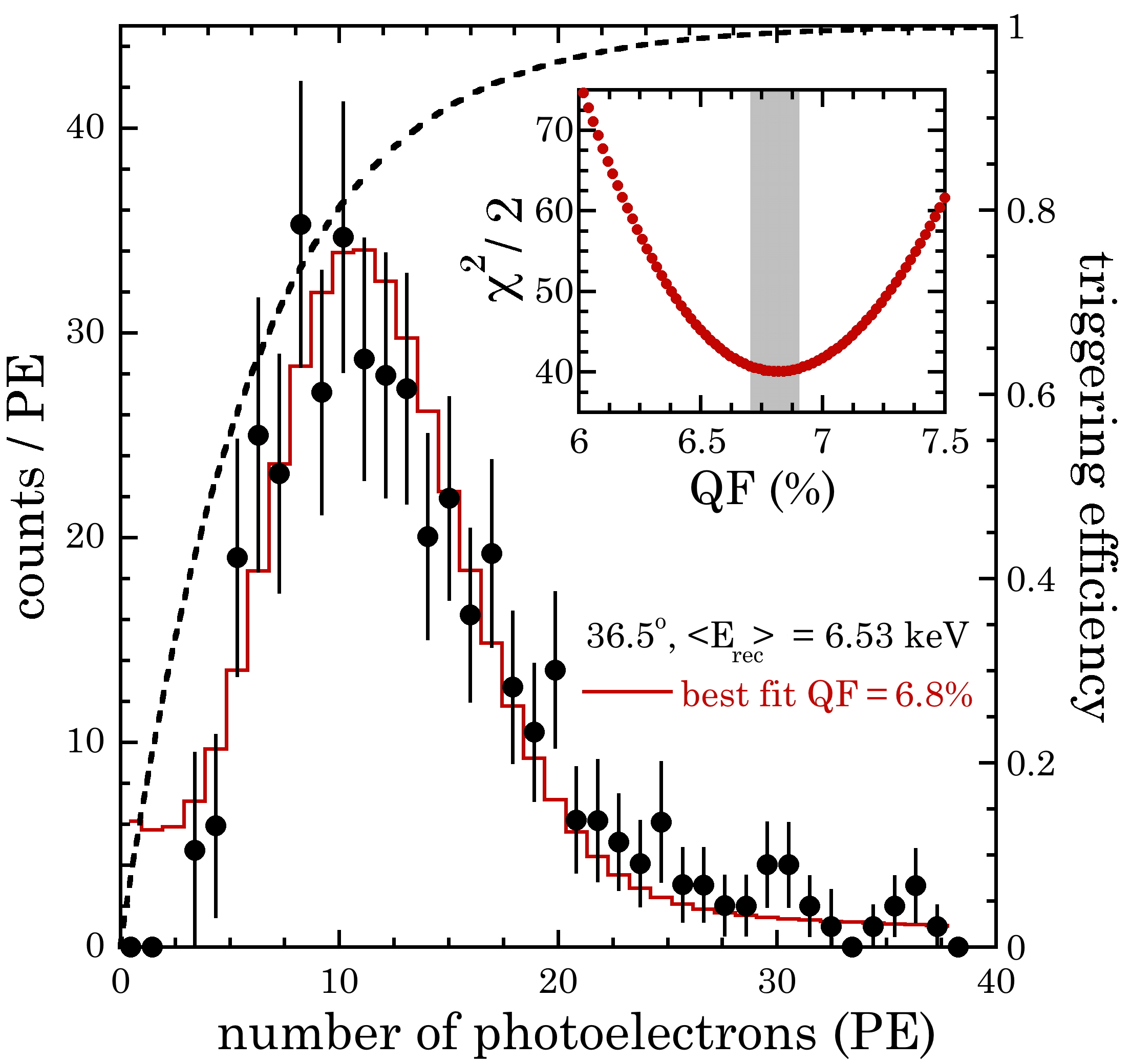}
    \caption{Comparison between light yield observed for $\sim$6.5 keV NRs at 108 K (datapoints), and its best-fit simulated prediction (histogram). The triggering efficiency, calculated as in \cite{setupNa,setupSNS}, is also shown, and is corrected for prior to data comparison with simulations. A band in the inset is the $\pm$1-$\sigma$ uncertainty in the best-fit QF.
    }
    \label{fig:bestfit}
\end{figure}

Also following \cite{csiqf}, we utilize an integrated rise-time (IRT) analysis \cite{irt1,irt2} to separate neutron- from gamma-induced events in the Bicron 501A backing detector. The resulting data quality is illustrated in Fig.\ \ref{fig:NScatter} following rejection of initially-dominant gamma contaminations. The modest background of random coincidences between CsI and backing detector is removed by subtraction of the energy spectrum of events within the 105-225 ns time range of Fig.\ \ref{fig:NScatter} from that for true coincidences concentrated within the 225-345 ns interval. The residual spectra of NR signals from elastic neutron scattering at 108 K are shown in Fig.\ \ref{fig:108K_layout}, for each of the six scattering angles explored. 

\begin{figure}
    \centering
    \includegraphics[width=.77\linewidth]{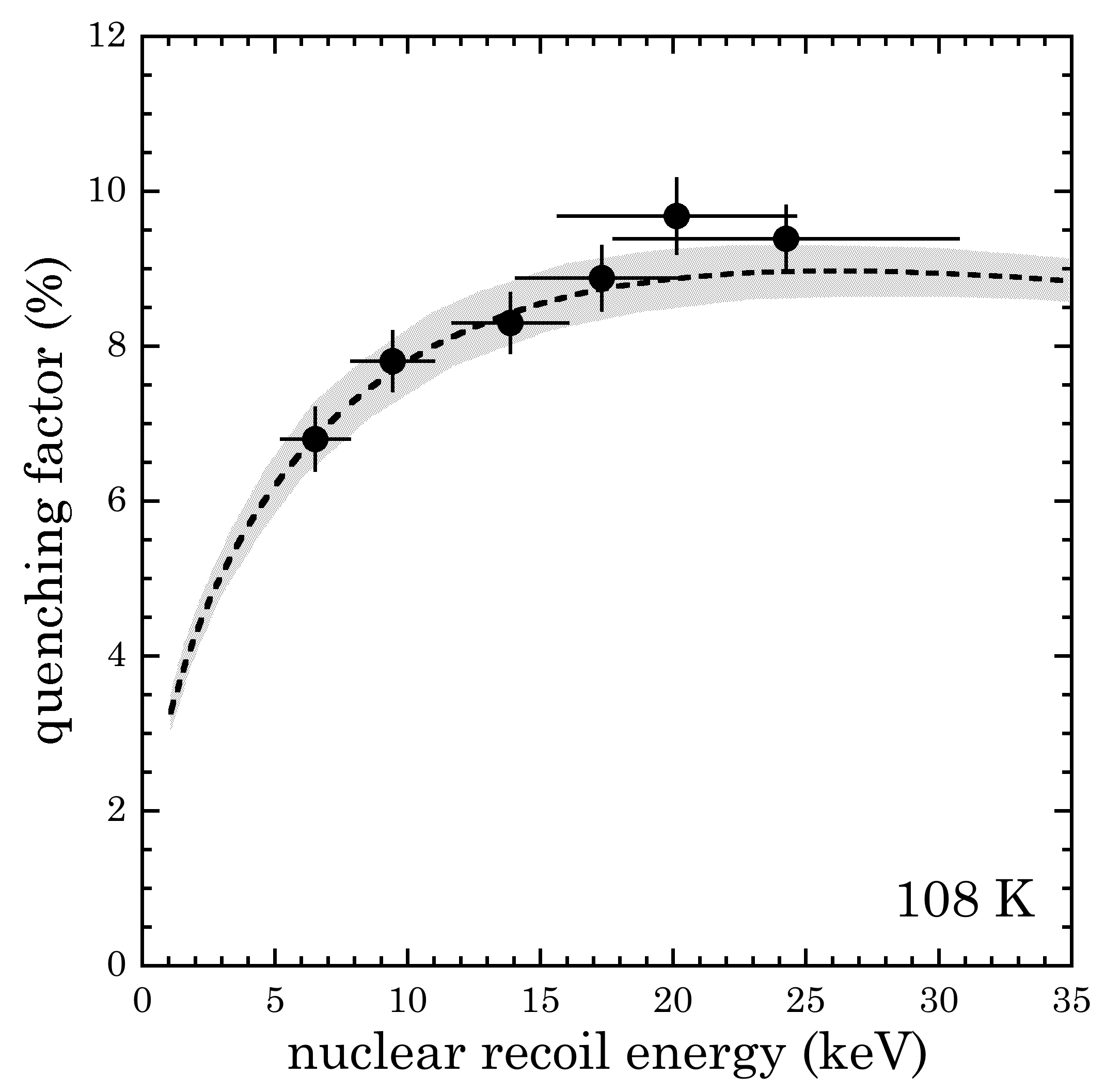}
    \caption{Quenching factor for low-energy nuclear recoils in undoped cryogenic CsI. The recoil energies probed span the CE$\nu$NS range of interest for CsI at a spallation source \cite{sciencepaper,ESS}. A dashed line shows the modified Birks model developed in \cite{csiqf} for 295 K CsI[Na], a grayed band its $\pm$1-$\sigma$ uncertainty.}
    \label{fig:QFvsE}
\end{figure}

The extraction of a best-fit QF is accomplished identically to \cite{csiqf}:  simulated energy depositions are translated to a corresponding number of photoelectrons (PE), accounting for Poisson smearing of PE statistics and the effect of the assumed QF. The obtained simulated distributions are then compared with the experimental residuals of Fig.\ \ref{fig:108K_layout}, with a log-likelihood analysis selecting the most adequate QF. Fig.\ \ref{fig:bestfit} illustrates this procedure for the lowest-energy NRs measured. The uncertainties in the QF values, manifested as vertical error bars in Fig.\ \ref{fig:QFvsE}, combine the 1-sigma log-likelihood error and the small dispersion in the  $^{241}$Am light yield. This energy reference was measured repeatedly during these runs. Horizontal error bars in Fig.\ \ref{fig:QFvsE} are akin to those in \cite{csiqf}, i.e., derived from the simulated spread in NR energies probed.

The totality of our QF  measurements for pure CsI at 108 K are reported in Fig.\ \ref{fig:QFvsE}. An excellent match to the modified Birks model developed for room-temperature CsI[Na] in \cite{csiqf} is noticeable. At least from the point of view of the adiabatic factor included in that model this agreement is not surprising: the band gap on which this adiabatic factor depends is not expected to change significantly from room-temperature to 108 K \cite{semi_bandgap,CsI_DebyeTemp}, an argument supported by observations in other cryogenic scintillators \cite{LiMoO_bandgapvsT}. However, this apparent constancy of the QF for NRs over the 108-295 K temperature range is in contrast with a reported factor of $\sim$7 increase with decreasing temperature in the QF for alpha particles, over the same range, for this material \cite{nadeau,clark}.

\begin{figure}
    \centering
    \includegraphics[width=.84\linewidth]{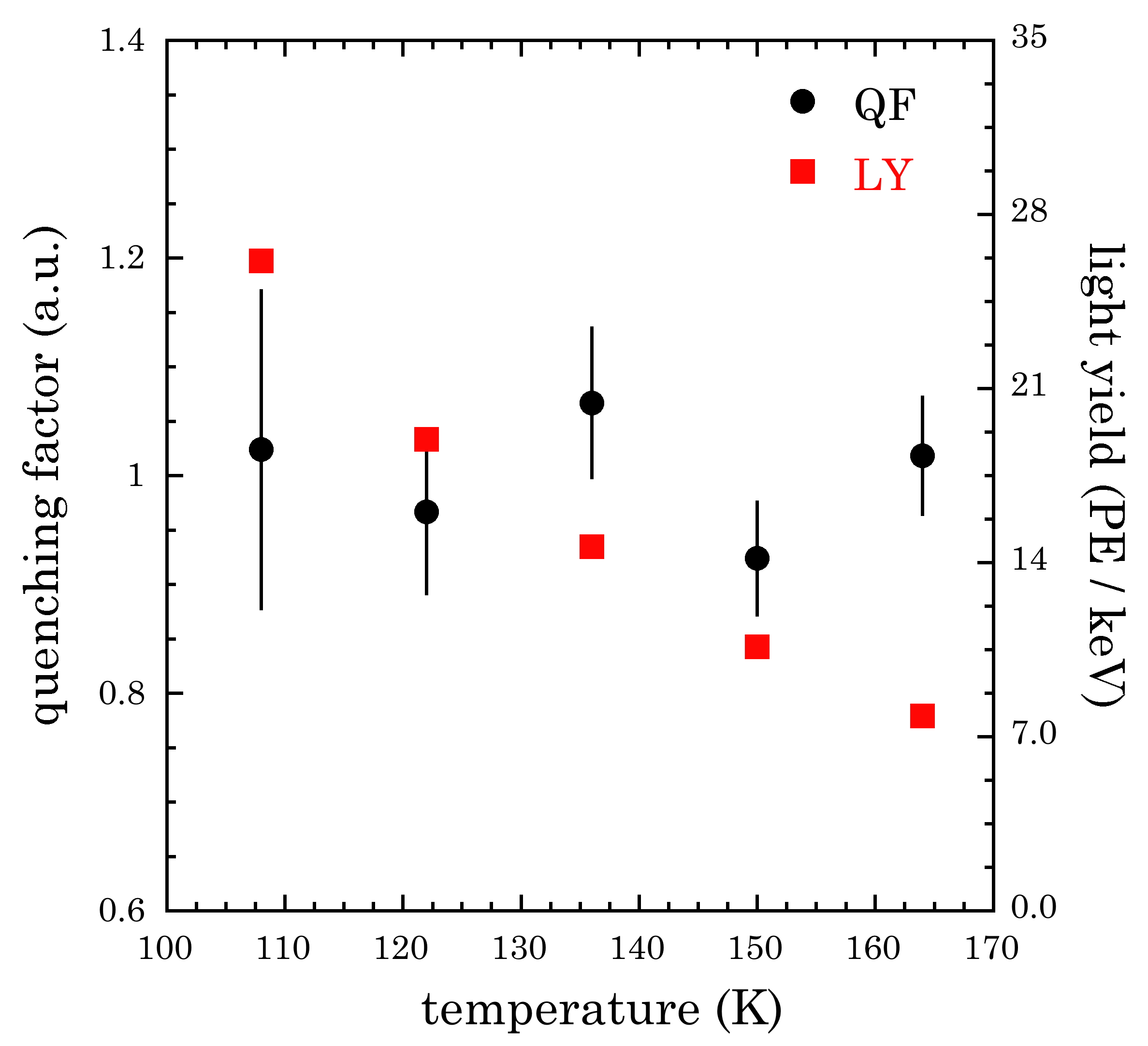}
    \caption{CsI quenching factor measurements as a function of temperature, normalized to their average, for the 56$^\circ$ scattering angle (13.9 keV NRs). No significant dependence of QF on temperature is observed. The $^{241}$Am light yield shown follows a trend of rapid change as in \cite{amsler,mik,woody,zhang} (error bars are encumbered by datapoints).  Extrapolated to 87 K, the lowest operating temperature of modern bialkali PMTs, the observed light yield triples that for room-temperature CsI[Na] during the first CE$\nu$NS observation \cite{sciencepaper,Scholz_2018}. }
    \label{fig:QFvsT}
\end{figure}

To confirm the observed independence of the QF on operating temperature, measurements at the 56$^\circ$ scattering angle (13.9 keV NR energy) were repeated for four additional temperatures, up to 165 K. The maximum temperature that could be explored was limited by the rapidly decreasing light yield. Fig.\ \ref{fig:QFvsT} shows the result of these measurements. As expected, no statistically-significant variation in the QF is visible, over a temperature range for which the overall light yield nevertheless more than tripled. 

\section{ancillary measurements} \label{resultsER}

\begin{figure}
    \centering
    \includegraphics[width=.8\linewidth]{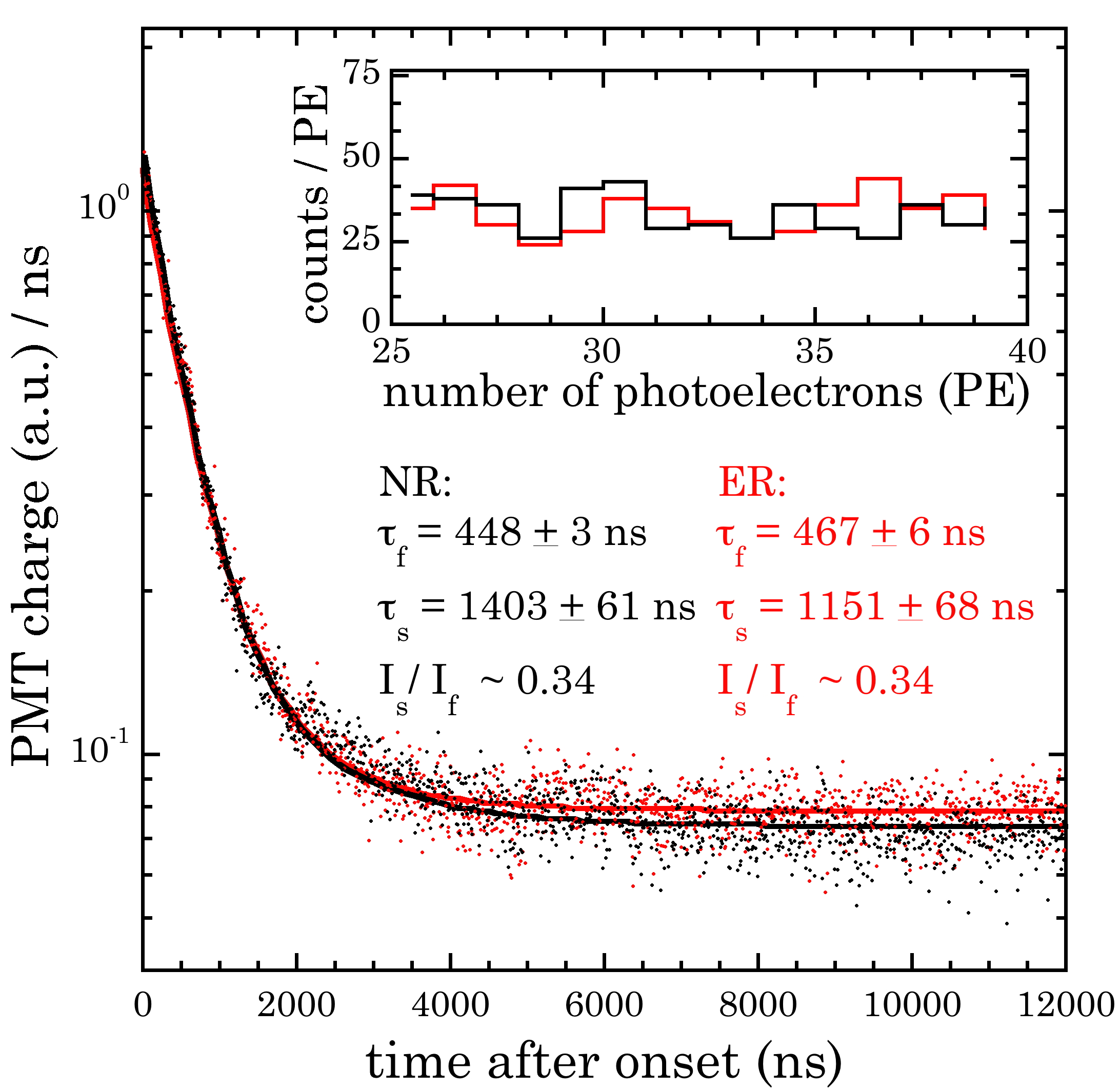}
    \caption{Decomposition of the scintillation decay time of pure CsI at 108 K into fast and slow components, for co-added ensembles of low-energy NRs and ERs (see text). One in ten waveform points is displayed, for clarity. For an unbiased ER-NR comparison, the PE distributions (inset) were chosen for similarity between both data sets. Best-fit slow (s) and fast (f) scintillation decay constants, and the ratio of PMT current in each decay component are shown.}
    \label{fig:decay}
\end{figure}

Small differences in the scintillation decay properties of ERs and NRs have been exploited in past experiments. Even when these are too subtle for event-by-event ER-NR discrimination they can still be applied to a large enough ensemble of events, statistically improving the sensitivity of a search for rare NR events \cite{decayPSD}.  To explore this possibility, a dedicated ER data set was collected containing Compton scatters from a collimated beam of $^{133}$Ba gammas impinging on the CsI crystal. Low-energy events were favored by triggering the DAQ on coincidences with a backing detector placed at a small angle with respect to the incoming beam \cite{Scholz_2018}. Five hundred events were selected from this data set, and the same number from available NR data, with the criterion that both groups should have similar distributions in the number of PE registered per event (Fig.\ \ref{fig:decay} inset, \cite{setupSNS}). This PE range selection corresponds to a NR energy of $\sim$15-25 keV. Waveforms within these two subsets were co-added, aligning all traces at the position of the first PE in each. The resulting artificial spike in PMT current at time-zero was removed from the analysis \cite{setupSNS}. The average ER and NR traces thus obtained, shown in Fig.\ \ref{fig:decay}, were fitted allowing for fast and slow scintillation decay components \cite{amsler,woody}. The PMT overshoot corrections for each data subset had identical decay components, to avoid introducing an artifact in the fits. This direct comparison between few-keV ER and NR events in CsI at 108 K shows only subtle differences, probably too difficult to exploit even for statistical ER-NR discrimination.

\begin{figure}
    \centering
    \includegraphics[width=.8\linewidth]{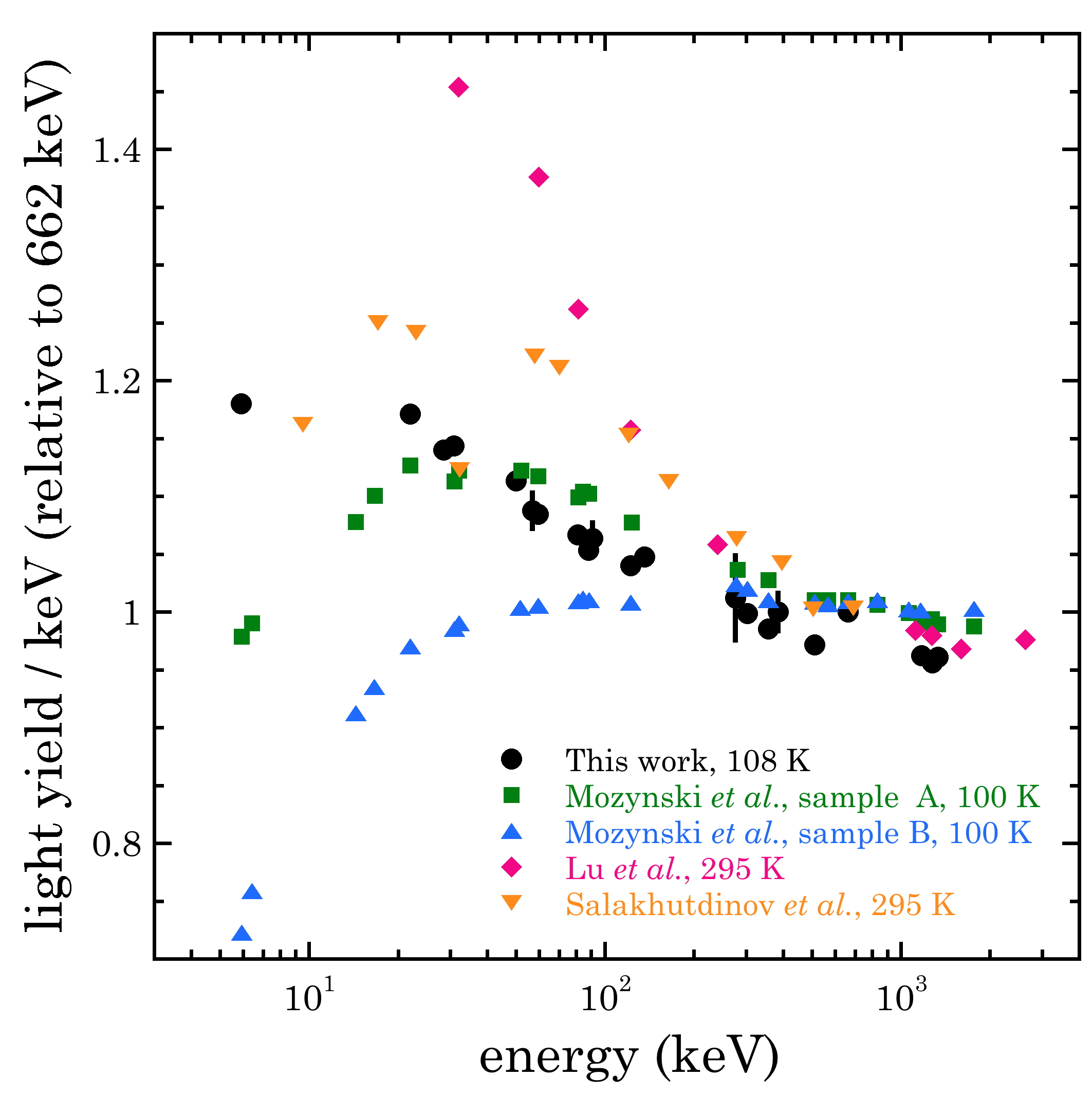}
    \caption{All known light yield proportionality data for ERs in undoped CsI at various temperatures, made relative to 662 keV \cite{mos1,Lu_CsIRmT,lotofscin}, including the present measurement.}
    \label{fig:nonprop}
\end{figure}

Separately, exposures to a variety of gamma-emitting radiosotopes were obtained in order to define the light yield proportionality of pure CsI at 108 K. A lowest-energy datapoint at 5.9 keV was acquired by placing an evaporated $^{55}$Fe source adjacent to the CsI crystal, in contact with its PTFE reflector. These results are displayed in Fig.\ \ref{fig:nonprop}, along with all other available similar data for this material \cite{mos1,Lu_CsIRmT,lotofscin}. Attempts have been made to understand the considerable dispersion in these results as a function of operating temperature and of CsI sample origin \cite{Lu_CsIRmT}. Our measurements using Amcrys/Proteus stock \cite{proteus} show a characteristic absence of reduction in light yield below $\sim$30 keV ER energy, seen in other datasets. The deviation from the assumption made in the definition of the QF that direct proportionality exists below 59.5 keV ER energy, is modest for our data. As emphasized in \cite{csiqf}, this assumption is in any case immaterial as long as the NR energy scale it defines is applied consistently to both QF calibrations and in the interpretation of physics runs. 

\begin{figure}
    \centering
    \includegraphics[width=.8\linewidth]{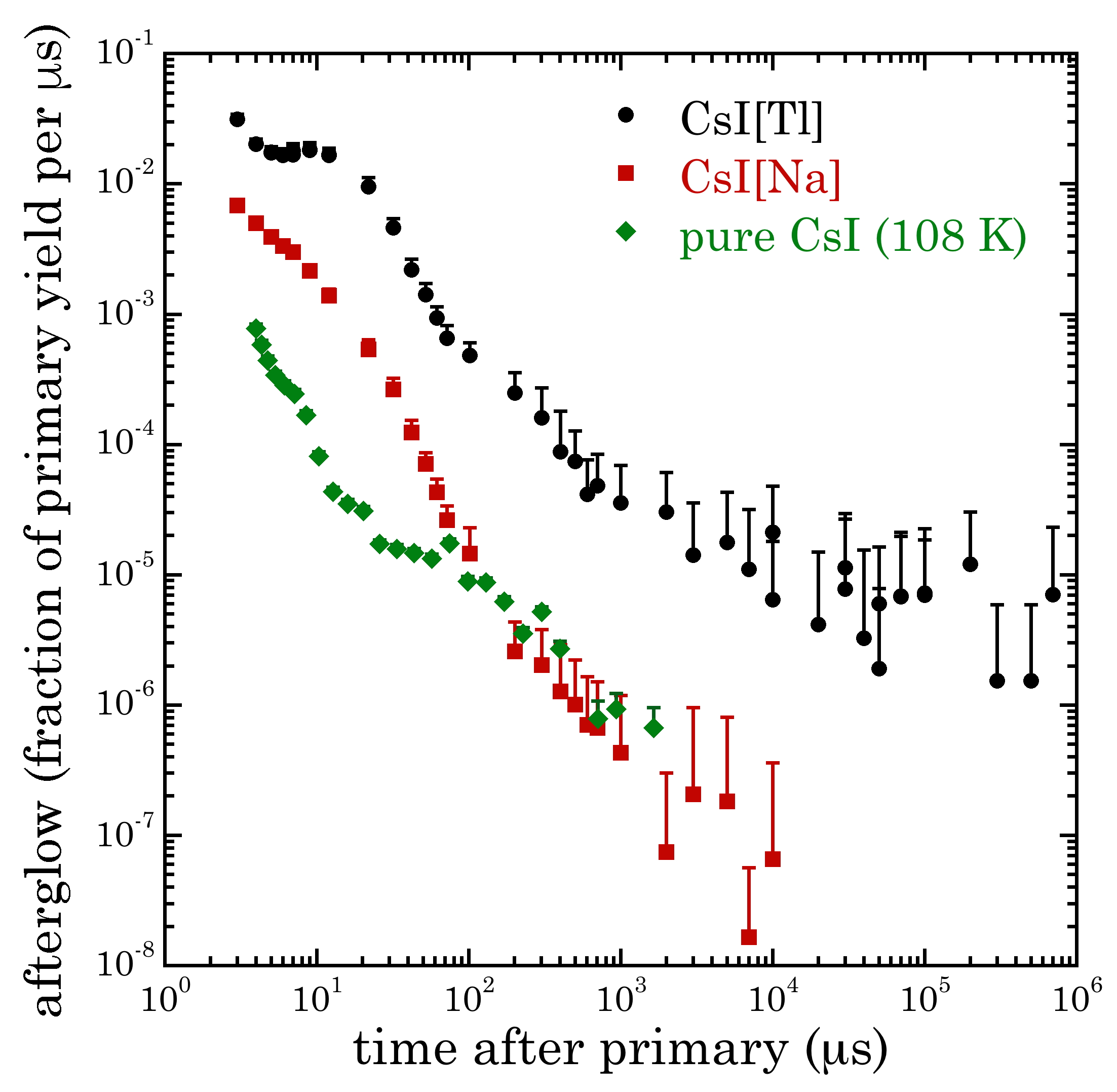}
    \caption{Afterglow in cryogenic pure CsI. The procedure to obtain these data follows our previous CsI[Na] and CsI[Tl] measurements, also shown \cite{setupSNS}. Each data point combines 500 measurements. This was 100 for doped material, leading to smaller present error bars, shown one-sided for clarity.}
    \label{fig:aftglw}
\end{figure}

The light yield previously demonstrated for pure CsI at  liquid nitrogen temperature ($\sim$80 K) is in the range of 80-125 scintillation photons per keV at a reference ER energy of 662 keV \cite{amsler,mos2,clark,liu}, displaying a dependence on CsI stock \cite{mos1,mu2_csi}. The presently measured yield is 26.13 $\pm$ 0.37 photoelectrons per keV at 108 K (Fig.\ \ref{fig:QFvsT}). Accounting for the 25$\%$ quantum efficiency of R8520-506 PMTs at the $\sim$340 nm emission characteristic of cryogenic pure CsI \cite{woody,amsler}, and a non-proportionality of 8.5$\%$ between 59.5 keV and 662 keV (Fig.\ \ref{fig:nonprop}), gives \mbox{96.3 $\pm$ 1.4} scintillation photons per keV at 108 K, based on our data. Following the temperature trends observed in \cite{amsler,mik,woody,zhang} and Fig.\ \ref{fig:QFvsT}, an additional $\sim$10\% increase is to be expected at the minimum 87 K operating temperature of present-day cryogenic bialkali PMTs. As emphasized in \cite{zhang,clark,ESS,liu}, this uncommonly-high yield is ideal for low-energy NR detection.

A final study was performed to quantify the afterglow (phosphorescence) of cryogenic pure CsI, thus far also an unknown quantity. These long-delayed few-PE emissions following a primary energy deposition can lead to a continuum of low-energy pulses that impede the identification of NRs, raising the effective threshold of the detector. The abatement of afterglow is of particular importance for CE$\nu$NS searches, performed without the benefit of a significant overburden, and therefore subject to frequent, energetic  primaries from cosmic-ray traversal. CsI[Na] was preferred over CsI[Tl] during the first CE$\nu$NS measurement for this reason \cite{setupSNS}: however, the removal of residual afterglow events still resulted in significant signal acceptance losses \cite{sciencepaper,Scholz_2018}. Fig.\ \ref{fig:aftglw} illustrates our results, following the same procedure as in \cite{setupSNS} (integration of afterglow over 1 $\mu$s periods following a $\sim$1.5 MeV primary). The substantial further inhibition of this process for cryogenic CsI is immediately noticeable to an experimenter upon simple inspection of oscilloscope traces: this bodes well for its use in the detection of faint scintillation signals.

\section{Conclusions} \label{Conclusion}



We have presented a first measurement of the quenching factor for low-energy nuclear recoils in undoped CsI at cryogenic temperature. Our results indicate that it is indistinguishable from that for room-temperature CsI[Na], further validating a physical response  model developed for that scintillator in \cite{csiqf}, and the modified interpretation of the first CE$\nu$NS measurement described in that same publication. The combination of a sizeable and by now well-understood NR quenching factor, negligible afterglow, and light yield in excess of 100 scintillation photons per keV defines an exceptionally promising material for WIMP and CE$\nu$NS detection. Specifically, when combined with high quantum-efficiency light sensors, cryogenic CsI can provide a sensitivity to $\sim$1 keV nuclear recoils, a new frontier for scintillating materials \cite{ESS}. In the context of the high neutrino flux expected from the upcoming European Spallation Source, a small array of cryogenic CsI scintillators can provide an unprecedented sensitivity to physics beyond the Standard Model via CE$\nu$NS studies \cite{ESS}.   

Additional work using a $^{88}$Y/Be photoneutron source and the QF measurement technique laid out in \cite{Scholz_GEconYBe,Collar_YBe,alvaro} is planned. This will probe nuclear recoils below 4.6 keV in CsI. A search for low-energy deviations from the model depicted in Fig.\ \ref{fig:QFvsE}, stemming from Migdal-like processes \cite{Migdal,yoni,tongyan}, should be possible using this unique material.\\

\section{Acknowledgements} \label{Acknowledgements}

We are indebted to P. Parkhurst at Proteus Inc. for many useful consultations. This work was supported by NSF awards PHY-1806722, PHY-1812702, and by the Kavli Institute for Cosmological Physics at the University of Chicago through an endowment from the Kavli Foundation and its founder Fred Kavli.

\bibliography{bibliography}

\begin{thebibliography}{39}%
\makeatletter
\providecommand \@ifxundefined [1]{%
 \@ifx{#1\undefined}
}%
\providecommand \@ifnum [1]{%
 \ifnum #1\expandafter \@firstoftwo
 \else \expandafter \@secondoftwo
 \fi
}%
\providecommand \@ifx [1]{%
 \ifx #1\expandafter \@firstoftwo
 \else \expandafter \@secondoftwo
 \fi
}%
\providecommand \natexlab [1]{#1}%
\providecommand \enquote  [1]{``#1''}%
\providecommand \bibnamefont  [1]{#1}%
\providecommand \bibfnamefont [1]{#1}%
\providecommand \citenamefont [1]{#1}%
\providecommand \href@noop [0]{\@secondoftwo}%
\providecommand \href [0]{\begingroup \@sanitize@url \@href}%
\providecommand \@href[1]{\@@startlink{#1}\@@href}%
\providecommand \@@href[1]{\endgroup#1\@@endlink}%
\providecommand \@sanitize@url [0]{\catcode `\\12\catcode `\$12\catcode
  `\&12\catcode `\#12\catcode `\^12\catcode `\_12\catcode `\%12\relax}%
\providecommand \@@startlink[1]{}%
\providecommand \@@endlink[0]{}%
\providecommand \url  [0]{\begingroup\@sanitize@url \@url }%
\providecommand \@url [1]{\endgroup\@href {#1}{\urlprefix }}%
\providecommand \urlprefix  [0]{URL }%
\providecommand \Eprint [0]{\href }%
\providecommand \doibase [0]{http://dx.doi.org/}%
\providecommand \selectlanguage [0]{\@gobble}%
\providecommand \bibinfo  [0]{\@secondoftwo}%
\providecommand \bibfield  [0]{\@secondoftwo}%
\providecommand \translation [1]{[#1]}%
\providecommand \BibitemOpen [0]{}%
\providecommand \bibitemStop [0]{}%
\providecommand \bibitemNoStop [0]{.\EOS\space}%
\providecommand \EOS [0]{\spacefactor3000\relax}%
\providecommand \BibitemShut  [1]{\csname bibitem#1\endcsname}%
\let\auto@bib@innerbib\@empty
\bibitem [{\citenamefont {Ahlen}\ \emph {et~al.}(1987)\citenamefont {Ahlen},
  \citenamefont {Avignone}, \citenamefont {Brodzinski}, \citenamefont
  {Drukier}, \citenamefont {Gelmini},\ and\ \citenamefont {Spergel}}]{wimp1}%
  \BibitemOpen
  \bibfield  {author} {\bibinfo {author} {\bibfnamefont {S.}~\bibnamefont
  {Ahlen}}, \bibinfo {author} {\bibfnamefont {F.}~\bibnamefont {Avignone}},
  \bibinfo {author} {\bibfnamefont {R.}~\bibnamefont {Brodzinski}}, \bibinfo
  {author} {\bibfnamefont {A.}~\bibnamefont {Drukier}}, \bibinfo {author}
  {\bibfnamefont {G.}~\bibnamefont {Gelmini}}, \ and\ \bibinfo {author}
  {\bibfnamefont {D.}~\bibnamefont {Spergel}},\ }\href {\doibase
  https://doi.org/10.1016/0370-2693(87)91581-4} {\bibfield  {journal} {\bibinfo
   {journal} {Phys. Lett. B}\ }\textbf {\bibinfo {volume} {195}},\ \bibinfo
  {pages} {603 } (\bibinfo {year} {1987})}\BibitemShut {NoStop}%
\bibitem [{\citenamefont {Freedman}(1974)}]{freedman}%
  \BibitemOpen
  \bibfield  {author} {\bibinfo {author} {\bibfnamefont {D.~Z.}\ \bibnamefont
  {Freedman}},\ }\href {\doibase 10.1103/PhysRevD.9.1389} {\bibfield  {journal}
  {\bibinfo  {journal} {Phys. Rev. D}\ }\textbf {\bibinfo {volume} {9}},\
  \bibinfo {pages} {1389} (\bibinfo {year} {1974})}\BibitemShut {NoStop}%
\bibitem [{\citenamefont {Akimov}\ \emph {et~al.}(2017)\citenamefont {Akimov}
  \emph {et~al.}}]{sciencepaper}%
  \BibitemOpen
  \bibfield  {author} {\bibinfo {author} {\bibfnamefont {D.}~\bibnamefont
  {Akimov}} \emph {et~al.},\ }\href {\doibase 10.1126/science.aao0990}
  {\bibfield  {journal} {\bibinfo  {journal} {Science}\ }\textbf {\bibinfo
  {volume} {357}},\ \bibinfo {pages} {1123} (\bibinfo {year}
  {2017})}\BibitemShut {NoStop}%
\bibitem [{\citenamefont {Scholz}(2017)}]{Scholz_2018}%
  \BibitemOpen
  \bibfield  {author} {\bibinfo {author} {\bibfnamefont {B.}~\bibnamefont
  {Scholz}},\ }\href@noop {} {Ph.D. thesis},\ \bibinfo  {school} {University of
  Chicago} (\bibinfo {year} {2017}),\ \Eprint {http://arxiv.org/abs/1904.01155}
  {arXiv:1904.01155} \BibitemShut {NoStop}%
\bibitem [{\citenamefont {Collar}\ \emph {et~al.}(2015)\citenamefont {Collar},
  \citenamefont {Fields}, \citenamefont {Hai}, \citenamefont {Hossbach},
  \citenamefont {Orrell}, \citenamefont {Overman}, \citenamefont {Perumpilly},\
  and\ \citenamefont {Scholz}}]{setupSNS}%
  \BibitemOpen
  \bibfield  {author} {\bibinfo {author} {\bibfnamefont {J.~I.}\ \bibnamefont
  {Collar}}, \bibinfo {author} {\bibfnamefont {N.~E.}\ \bibnamefont {Fields}},
  \bibinfo {author} {\bibfnamefont {M.}~\bibnamefont {Hai}}, \bibinfo {author}
  {\bibfnamefont {T.~W.}\ \bibnamefont {Hossbach}}, \bibinfo {author}
  {\bibfnamefont {J.~L.}\ \bibnamefont {Orrell}}, \bibinfo {author}
  {\bibfnamefont {C.~T.}\ \bibnamefont {Overman}}, \bibinfo {author}
  {\bibfnamefont {G.}~\bibnamefont {Perumpilly}}, \ and\ \bibinfo {author}
  {\bibfnamefont {B.}~\bibnamefont {Scholz}},\ }\href {\doibase
  https://doi.org/10.1016/j.nima.2014.11.037} {\bibfield  {journal} {\bibinfo
  {journal} {Nucl. Instr. Meth. A}\ }\textbf {\bibinfo {volume} {773}},\
  \bibinfo {pages} {56 } (\bibinfo {year} {2015})}\BibitemShut {NoStop}%
\bibitem [{\citenamefont {Fields}(2014)}]{nicolethesis}%
  \BibitemOpen
  \bibfield  {author} {\bibinfo {author} {\bibfnamefont {N.}~\bibnamefont
  {Fields}},\ }\href@noop {} {Ph.D. thesis},\ \bibinfo  {school} {University of
  Chicago} (\bibinfo {year} {2014})\BibitemShut {NoStop}%
\bibitem [{\citenamefont {Collar}\ \emph {et~al.}(2019)\citenamefont {Collar},
  \citenamefont {Kavner},\ and\ \citenamefont {Lewis}}]{csiqf}%
  \BibitemOpen
  \bibfield  {author} {\bibinfo {author} {\bibfnamefont {J.~I.}\ \bibnamefont
  {Collar}}, \bibinfo {author} {\bibfnamefont {A.~R.~L.}\ \bibnamefont
  {Kavner}}, \ and\ \bibinfo {author} {\bibfnamefont {C.~M.}\ \bibnamefont
  {Lewis}},\ }\href {\doibase 10.1103/PhysRevD.100.033003} {\bibfield
  {journal} {\bibinfo  {journal} {Phys. Rev. D}\ }\textbf {\bibinfo {volume}
  {100}},\ \bibinfo {pages} {033003} (\bibinfo {year} {2019})}\BibitemShut
  {NoStop}%
\bibitem [{\citenamefont {Amsler}\ \emph {et~al.}(2002)\citenamefont {Amsler},
  \citenamefont {Grogler}, \citenamefont {Joffrain}, \citenamefont {Lindelof},
  \citenamefont {Marchesotti}, \citenamefont {Niederberger}, \citenamefont
  {Pruys}, \citenamefont {Regenfus}, \citenamefont {Riedler},\ and\
  \citenamefont {Rotondi}}]{amsler}%
  \BibitemOpen
  \bibfield  {author} {\bibinfo {author} {\bibfnamefont {C.}~\bibnamefont
  {Amsler}}, \bibinfo {author} {\bibfnamefont {D.}~\bibnamefont {Grogler}},
  \bibinfo {author} {\bibfnamefont {W.}~\bibnamefont {Joffrain}}, \bibinfo
  {author} {\bibfnamefont {D.}~\bibnamefont {Lindelof}}, \bibinfo {author}
  {\bibfnamefont {M.}~\bibnamefont {Marchesotti}}, \bibinfo {author}
  {\bibfnamefont {P.}~\bibnamefont {Niederberger}}, \bibinfo {author}
  {\bibfnamefont {H.}~\bibnamefont {Pruys}}, \bibinfo {author} {\bibfnamefont
  {C.}~\bibnamefont {Regenfus}}, \bibinfo {author} {\bibfnamefont
  {P.}~\bibnamefont {Riedler}}, \ and\ \bibinfo {author} {\bibfnamefont
  {A.}~\bibnamefont {Rotondi}},\ }\href {\doibase
  https://doi.org/10.1016/S0168-9002(01)01239-6} {\bibfield  {journal}
  {\bibinfo  {journal} {Nucl. Instr. Meth. A}\ }\textbf {\bibinfo {volume}
  {480}},\ \bibinfo {pages} {494 } (\bibinfo {year} {2002})}\BibitemShut
  {NoStop}%
\bibitem [{\citenamefont {Moszynski}\ \emph {et~al.}(2005)\citenamefont
  {Moszynski}, \citenamefont {Balcerzyk}, \citenamefont {Czarnacki},
  \citenamefont {Kapusta}, \citenamefont {Klamra}, \citenamefont {Schotanus},
  \citenamefont {Syntfeld}, \citenamefont {Szawcowski},\ and\ \citenamefont
  {Kozlov}}]{mos1}%
  \BibitemOpen
  \bibfield  {author} {\bibinfo {author} {\bibfnamefont {M.}~\bibnamefont
  {Moszynski}}, \bibinfo {author} {\bibfnamefont {M.}~\bibnamefont
  {Balcerzyk}}, \bibinfo {author} {\bibfnamefont {W.}~\bibnamefont
  {Czarnacki}}, \bibinfo {author} {\bibfnamefont {M.}~\bibnamefont {Kapusta}},
  \bibinfo {author} {\bibfnamefont {W.}~\bibnamefont {Klamra}}, \bibinfo
  {author} {\bibfnamefont {P.}~\bibnamefont {Schotanus}}, \bibinfo {author}
  {\bibfnamefont {A.}~\bibnamefont {Syntfeld}}, \bibinfo {author}
  {\bibfnamefont {M.}~\bibnamefont {Szawcowski}}, \ and\ \bibinfo {author}
  {\bibfnamefont {V.}~\bibnamefont {Kozlov}},\ }\href {\doibase
  10.1016/j.nima.2004.08.043} {\bibfield  {journal} {\bibinfo  {journal} {Nucl.
  Instr. Meth. A}\ }\textbf {\bibinfo {volume} {537}},\ \bibinfo {pages} {357}
  (\bibinfo {year} {2005})}\BibitemShut {NoStop}%
\bibitem [{\citenamefont {Moszynski}\ \emph {et~al.}(2003)\citenamefont
  {Moszynski}, \citenamefont {Czarnacki}, \citenamefont {Klamra}, \citenamefont
  {Szawlowski}, \citenamefont {Schotanus},\ and\ \citenamefont
  {Kapusta}}]{mos2}%
  \BibitemOpen
  \bibfield  {author} {\bibinfo {author} {\bibfnamefont {M.}~\bibnamefont
  {Moszynski}}, \bibinfo {author} {\bibfnamefont {W.}~\bibnamefont
  {Czarnacki}}, \bibinfo {author} {\bibfnamefont {W.}~\bibnamefont {Klamra}},
  \bibinfo {author} {\bibfnamefont {M.}~\bibnamefont {Szawlowski}}, \bibinfo
  {author} {\bibfnamefont {P.}~\bibnamefont {Schotanus}}, \ and\ \bibinfo
  {author} {\bibfnamefont {M.}~\bibnamefont {Kapusta}},\ }\href {\doibase
  https://doi.org/10.1016/S0168-9002(03)00785-X} {\bibfield  {journal}
  {\bibinfo  {journal} {Nucl. Instr. Meth. A}\ }\textbf {\bibinfo {volume}
  {504}},\ \bibinfo {pages} {307 } (\bibinfo {year} {2003})}\BibitemShut
  {NoStop}%
\bibitem [{\citenamefont {Nadeau}(2015)}]{nadeau}%
  \BibitemOpen
  \bibfield  {author} {\bibinfo {author} {\bibfnamefont {P.}~\bibnamefont
  {Nadeau}},\ }\href@noop {} {Ph.D. thesis},\ \bibinfo  {school} {Queen's
  University} (\bibinfo {year} {2015})\BibitemShut {NoStop}%
\bibitem [{\citenamefont {Clark}\ \emph {et~al.}(2018)\citenamefont {Clark},
  \citenamefont {Nadeau}, \citenamefont {Hills}, \citenamefont {Dujardin},\
  and\ \citenamefont {Stefano}}]{clark}%
  \BibitemOpen
  \bibfield  {author} {\bibinfo {author} {\bibfnamefont {M.}~\bibnamefont
  {Clark}}, \bibinfo {author} {\bibfnamefont {P.}~\bibnamefont {Nadeau}},
  \bibinfo {author} {\bibfnamefont {S.}~\bibnamefont {Hills}}, \bibinfo
  {author} {\bibfnamefont {C.}~\bibnamefont {Dujardin}}, \ and\ \bibinfo
  {author} {\bibfnamefont {P.~D.}\ \bibnamefont {Stefano}},\ }\href {\doibase
  https://doi.org/10.1016/j.nima.2018.05.066} {\bibfield  {journal} {\bibinfo
  {journal} {Nucl. Instr. Meth. A}\ }\textbf {\bibinfo {volume} {901}},\
  \bibinfo {pages} {6 } (\bibinfo {year} {2018})}\BibitemShut {NoStop}%
\bibitem [{\citenamefont {Liu}\ \emph {et~al.}(2016)\citenamefont {Liu},
  \citenamefont {Yamashita},\ and\ \citenamefont {Soma}}]{liu}%
  \BibitemOpen
  \bibfield  {author} {\bibinfo {author} {\bibfnamefont {J.}~\bibnamefont
  {Liu}}, \bibinfo {author} {\bibfnamefont {M.}~\bibnamefont {Yamashita}}, \
  and\ \bibinfo {author} {\bibfnamefont {A.}~\bibnamefont {Soma}},\ }\href
  {\doibase 10.1088/1748-0221/11/10/p10003} {\bibfield  {journal} {\bibinfo
  {journal} {J. Instrum.}\ }\textbf {\bibinfo {volume} {11}},\ \bibinfo {pages}
  {P10003} (\bibinfo {year} {2016})}\BibitemShut {NoStop}%
\bibitem [{\citenamefont {{Woody}}\ \emph {et~al.}(1990)\citenamefont
  {{Woody}}, \citenamefont {{Levy}}, \citenamefont {{Kierstead}}, \citenamefont
  {{Skwarnicki}}, \citenamefont {{Sobolewski}}, \citenamefont {{Goldberg}},
  \citenamefont {{Horwitz}}, \citenamefont {{Souder}},\ and\ \citenamefont
  {{Anderson}}}]{woody}%
  \BibitemOpen
  \bibfield  {author} {\bibinfo {author} {\bibfnamefont {C.~L.}\ \bibnamefont
  {{Woody}}}, \bibinfo {author} {\bibfnamefont {P.~W.}\ \bibnamefont {{Levy}}},
  \bibinfo {author} {\bibfnamefont {J.~A.}\ \bibnamefont {{Kierstead}}},
  \bibinfo {author} {\bibfnamefont {T.}~\bibnamefont {{Skwarnicki}}}, \bibinfo
  {author} {\bibfnamefont {Z.}~\bibnamefont {{Sobolewski}}}, \bibinfo {author}
  {\bibfnamefont {M.}~\bibnamefont {{Goldberg}}}, \bibinfo {author}
  {\bibfnamefont {N.}~\bibnamefont {{Horwitz}}}, \bibinfo {author}
  {\bibfnamefont {P.}~\bibnamefont {{Souder}}}, \ and\ \bibinfo {author}
  {\bibfnamefont {D.~F.}\ \bibnamefont {{Anderson}}},\ }\href {\doibase
  10.1109/23.106667} {\bibfield  {journal} {\bibinfo  {journal} {IEEE Trans.
  Nucl. Sci.}\ }\textbf {\bibinfo {volume} {37}},\ \bibinfo {pages} {492}
  (\bibinfo {year} {1990})}\BibitemShut {NoStop}%
\bibitem [{\citenamefont {Zhang}\ \emph {et~al.}(2018)\citenamefont {Zhang},
  \citenamefont {Sun}, \citenamefont {Lu},\ and\ \citenamefont {Lu}}]{zhang}%
  \BibitemOpen
  \bibfield  {author} {\bibinfo {author} {\bibfnamefont {X.}~\bibnamefont
  {Zhang}}, \bibinfo {author} {\bibfnamefont {X.}~\bibnamefont {Sun}}, \bibinfo
  {author} {\bibfnamefont {J.}~\bibnamefont {Lu}}, \ and\ \bibinfo {author}
  {\bibfnamefont {P.}~\bibnamefont {Lu}},\ }\href {\doibase
  10.1007/s41605-018-0039-1} {\bibfield  {journal} {\bibinfo  {journal}
  {Radiat. Detect. Technol. Methods}\ }\textbf {\bibinfo {volume} {2}},\
  \bibinfo {pages} {15} (\bibinfo {year} {2018})}\BibitemShut {NoStop}%
\bibitem [{\citenamefont {Mikhailik}\ \emph {et~al.}(2015)\citenamefont
  {Mikhailik}, \citenamefont {Kapustyanyk}, \citenamefont {Tsybulskyi},
  \citenamefont {Rudyk},\ and\ \citenamefont {Kraus}}]{mik}%
  \BibitemOpen
  \bibfield  {author} {\bibinfo {author} {\bibfnamefont {V.~B.}\ \bibnamefont
  {Mikhailik}}, \bibinfo {author} {\bibfnamefont {V.}~\bibnamefont
  {Kapustyanyk}}, \bibinfo {author} {\bibfnamefont {V.}~\bibnamefont
  {Tsybulskyi}}, \bibinfo {author} {\bibfnamefont {V.}~\bibnamefont {Rudyk}}, \
  and\ \bibinfo {author} {\bibfnamefont {H.}~\bibnamefont {Kraus}},\ }\href
  {\doibase 10.1002/pssb.201451464} {\bibfield  {journal} {\bibinfo  {journal}
  {Phys. Status Solidi (b)}\ }\textbf {\bibinfo {volume} {252}},\ \bibinfo
  {pages} {804} (\bibinfo {year} {2015})}\BibitemShut {NoStop}%
\bibitem [{\citenamefont {Baxter}\ \emph {et~al.}(2020)\citenamefont {Baxter},
  \citenamefont {Collar}, \citenamefont {Coloma}, \citenamefont {Dahl},
  \citenamefont {Esteban}, \citenamefont {Ferrario}, \citenamefont
  {Gomez-Cadenas}, \citenamefont {Gonzalez-Garcia}, \citenamefont {Kavner},
  \citenamefont {Lewis}, \citenamefont {Monrabal}, \citenamefont {Mu\~{n}oz
  Vidal}, \citenamefont {Privitera}, \citenamefont {Ramanathan},\ and\
  \citenamefont {Renner}}]{ESS}%
  \BibitemOpen
  \bibfield  {author} {\bibinfo {author} {\bibfnamefont {D.}~\bibnamefont
  {Baxter}}, \bibinfo {author} {\bibfnamefont {J.~I.}\ \bibnamefont {Collar}},
  \bibinfo {author} {\bibfnamefont {P.}~\bibnamefont {Coloma}}, \bibinfo
  {author} {\bibfnamefont {C.~E.}\ \bibnamefont {Dahl}}, \bibinfo {author}
  {\bibfnamefont {I.}~\bibnamefont {Esteban}}, \bibinfo {author} {\bibfnamefont
  {P.}~\bibnamefont {Ferrario}}, \bibinfo {author} {\bibfnamefont {J.~J.}\
  \bibnamefont {Gomez-Cadenas}}, \bibinfo {author} {\bibfnamefont
  {M.}~\bibnamefont {Gonzalez-Garcia}}, \bibinfo {author} {\bibfnamefont
  {A.~R.~L.}\ \bibnamefont {Kavner}}, \bibinfo {author} {\bibfnamefont {C.~M.}\
  \bibnamefont {Lewis}}, \bibinfo {author} {\bibfnamefont {F.}~\bibnamefont
  {Monrabal}}, \bibinfo {author} {\bibfnamefont {J.}~\bibnamefont {Mu\~{n}oz
  Vidal}}, \bibinfo {author} {\bibfnamefont {P.}~\bibnamefont {Privitera}},
  \bibinfo {author} {\bibfnamefont {K.}~\bibnamefont {Ramanathan}}, \ and\
  \bibinfo {author} {\bibfnamefont {J.}~\bibnamefont {Renner}},\ }\href@noop {}
  {\bibfield  {journal} {\bibinfo  {journal} {JHEP}\ }\textbf {\bibinfo
  {volume} {2020}},\ \bibinfo {pages} {1} (\bibinfo {year} {2020})}\BibitemShut
  {NoStop}%
\bibitem [{\citenamefont {Collar}(2013{\natexlab{a}})}]{setupNa}%
  \BibitemOpen
  \bibfield  {author} {\bibinfo {author} {\bibfnamefont {J.~I.}\ \bibnamefont
  {Collar}},\ }\href {\doibase 10.1103/PhysRevC.88.035806} {\bibfield
  {journal} {\bibinfo  {journal} {Phys. Rev. C}\ }\textbf {\bibinfo {volume}
  {88}},\ \bibinfo {pages} {035806} (\bibinfo {year}
  {2013}{\natexlab{a}})}\BibitemShut {NoStop}%
\bibitem [{pro()}]{proteus}%
  \BibitemOpen
  \href@noop {} {}\bibinfo {note} {Amcrys stock, Czochralski grown in dedicated
  furnaces for trace-level dopant concentration. Procured from Proteus Inc.,
  Chagrin Falls, Ohio 44022, USA.}\BibitemShut {Stop}%
\bibitem [{\citenamefont {Pozzi}\ \emph {et~al.}(2003)\citenamefont {Pozzi},
  \citenamefont {Padovani},\ and\ \citenamefont {Marseguerra}}]{MCNP-PoliMi}%
  \BibitemOpen
  \bibfield  {author} {\bibinfo {author} {\bibfnamefont {S.~A.}\ \bibnamefont
  {Pozzi}}, \bibinfo {author} {\bibfnamefont {E.}~\bibnamefont {Padovani}}, \
  and\ \bibinfo {author} {\bibfnamefont {M.}~\bibnamefont {Marseguerra}},\
  }\href {\doibase https://doi.org/10.1016/j.nima.2003.06.012} {\bibfield
  {journal} {\bibinfo  {journal} {Nucl. Instr. Meth. A}\ }\textbf {\bibinfo
  {volume} {513}},\ \bibinfo {pages} {550 } (\bibinfo {year}
  {2003})}\BibitemShut {NoStop}%
\bibitem [{\citenamefont {Wright}\ and\ \citenamefont
  {Wright}(2017)}]{wright2017pmt}%
  \BibitemOpen
  \bibfield  {author} {\bibinfo {author} {\bibfnamefont {A.}~\bibnamefont
  {Wright}}\ and\ \bibinfo {author} {\bibfnamefont {T.}~\bibnamefont
  {Wright}},\ }\href {https://books.google.com/books?id=AqLAuQEACAAJ} {\emph
  {\bibinfo {title} {The Photomultiplier Handbook}}}\ (\bibinfo  {publisher}
  {Oxford University Press},\ \bibinfo {year} {2017})\BibitemShut {NoStop}%
\bibitem [{\citenamefont {Acciarri}\ \emph {et~al.}(2017)\citenamefont
  {Acciarri} \emph {et~al.}}]{boon}%
  \BibitemOpen
  \bibfield  {author} {\bibinfo {author} {\bibfnamefont {R.}~\bibnamefont
  {Acciarri}} \emph {et~al.},\ }\href {\doibase 10.1088/1748-0221/12/08/p08003}
  {\bibfield  {journal} {\bibinfo  {journal} {J. Instrum.}\ }\textbf {\bibinfo
  {volume} {12}},\ \bibinfo {pages} {P08003} (\bibinfo {year}
  {2017})}\BibitemShut {NoStop}%
\bibitem [{\citenamefont {Abe}\ \emph {et~al.}(2012)\citenamefont {Abe} \emph
  {et~al.}}]{chooz}%
  \BibitemOpen
  \bibfield  {author} {\bibinfo {author} {\bibfnamefont {Y.}~\bibnamefont
  {Abe}} \emph {et~al.} (\bibinfo {collaboration} {Double Chooz
  Collaboration}),\ }\href {\doibase 10.1103/PhysRevD.86.052008} {\bibfield
  {journal} {\bibinfo  {journal} {Phys. Rev. D}\ }\textbf {\bibinfo {volume}
  {86}},\ \bibinfo {pages} {052008} (\bibinfo {year} {2012})}\BibitemShut
  {NoStop}%
\bibitem [{\citenamefont {Zhang}\ \emph {et~al.}(2019)\citenamefont {Zhang},
  \citenamefont {Wang}, \citenamefont {Zhang}, \citenamefont {Huang},
  \citenamefont {Luo}, \citenamefont {Zhang}, \citenamefont {Zhang},
  \citenamefont {Xu}, \citenamefont {Liu}, \citenamefont {Heng}, \citenamefont
  {Yang}, \citenamefont {Jiang}, \citenamefont {Li}, \citenamefont {Ye},\ and\
  \citenamefont {Chen}}]{juno}%
  \BibitemOpen
  \bibfield  {author} {\bibinfo {author} {\bibfnamefont {H.}~\bibnamefont
  {Zhang}}, \bibinfo {author} {\bibfnamefont {Z.}~\bibnamefont {Wang}},
  \bibinfo {author} {\bibfnamefont {Y.}~\bibnamefont {Zhang}}, \bibinfo
  {author} {\bibfnamefont {Y.}~\bibnamefont {Huang}}, \bibinfo {author}
  {\bibfnamefont {F.}~\bibnamefont {Luo}}, \bibinfo {author} {\bibfnamefont
  {P.}~\bibnamefont {Zhang}}, \bibinfo {author} {\bibfnamefont
  {C.}~\bibnamefont {Zhang}}, \bibinfo {author} {\bibfnamefont
  {M.}~\bibnamefont {Xu}}, \bibinfo {author} {\bibfnamefont {J.}~\bibnamefont
  {Liu}}, \bibinfo {author} {\bibfnamefont {Y.}~\bibnamefont {Heng}}, \bibinfo
  {author} {\bibfnamefont {C.}~\bibnamefont {Yang}}, \bibinfo {author}
  {\bibfnamefont {X.}~\bibnamefont {Jiang}}, \bibinfo {author} {\bibfnamefont
  {F.}~\bibnamefont {Li}}, \bibinfo {author} {\bibfnamefont {M.}~\bibnamefont
  {Ye}}, \ and\ \bibinfo {author} {\bibfnamefont {H.}~\bibnamefont {Chen}},\
  }\href {\doibase 10.1088/1748-0221/14/08/t08002} {\bibfield  {journal}
  {\bibinfo  {journal} {J. Instrum.}\ }\textbf {\bibinfo {volume} {14}},\
  \bibinfo {pages} {T08002} (\bibinfo {year} {2019})}\BibitemShut {NoStop}%
\bibitem [{\citenamefont {Luo}\ \emph {et~al.}(2014)\citenamefont {Luo} \emph
  {et~al.}}]{irt1}%
  \BibitemOpen
  \bibfield  {author} {\bibinfo {author} {\bibfnamefont {X.~L.}\ \bibnamefont
  {Luo}} \emph {et~al.},\ }\href {\doibase
  https://doi.org/10.1016/j.nima.2014.08.023} {\bibfield  {journal} {\bibinfo
  {journal} {Nucl. Instr. Meth. A}\ }\textbf {\bibinfo {volume} {767}},\
  \bibinfo {pages} {83 } (\bibinfo {year} {2014})}\BibitemShut {NoStop}%
\bibitem [{\citenamefont {Ronchi}\ \emph {et~al.}(2009)\citenamefont {Ronchi},
  \citenamefont {Soderstrom}, \citenamefont {Nyberg}, \citenamefont {Sunden},
  \citenamefont {Conroy}, \citenamefont {Ericsson}, \citenamefont {Hellesen},
  \citenamefont {Johnson},\ and\ \citenamefont {Weiszflog}}]{irt2}%
  \BibitemOpen
  \bibfield  {author} {\bibinfo {author} {\bibfnamefont {E.}~\bibnamefont
  {Ronchi}}, \bibinfo {author} {\bibfnamefont {P.-A.}\ \bibnamefont
  {Soderstrom}}, \bibinfo {author} {\bibfnamefont {J.}~\bibnamefont {Nyberg}},
  \bibinfo {author} {\bibfnamefont {E.~A.}\ \bibnamefont {Sunden}}, \bibinfo
  {author} {\bibfnamefont {S.}~\bibnamefont {Conroy}}, \bibinfo {author}
  {\bibfnamefont {G.}~\bibnamefont {Ericsson}}, \bibinfo {author}
  {\bibfnamefont {C.}~\bibnamefont {Hellesen}}, \bibinfo {author}
  {\bibfnamefont {M.~G.}\ \bibnamefont {Johnson}}, \ and\ \bibinfo {author}
  {\bibfnamefont {M.}~\bibnamefont {Weiszflog}},\ }\href {\doibase
  https://doi.org/10.1016/j.nima.2009.08.064} {\bibfield  {journal} {\bibinfo
  {journal} {Nucl. Instr. Meth. A}\ }\textbf {\bibinfo {volume} {610}},\
  \bibinfo {pages} {534 } (\bibinfo {year} {2009})}\BibitemShut {NoStop}%
\bibitem [{\citenamefont {Varshni}(1967)}]{semi_bandgap}%
  \BibitemOpen
  \bibfield  {author} {\bibinfo {author} {\bibfnamefont {Y.}~\bibnamefont
  {Varshni}},\ }\href {\doibase https://doi.org/10.1016/0031-8914(67)90062-6}
  {\bibfield  {journal} {\bibinfo  {journal} {Physica}\ }\textbf {\bibinfo
  {volume} {34}},\ \bibinfo {pages} {149 } (\bibinfo {year}
  {1967})}\BibitemShut {NoStop}%
\bibitem [{\citenamefont {Karo}\ and\ \citenamefont
  {Hardy}(1968)}]{CsI_DebyeTemp}%
  \BibitemOpen
  \bibfield  {author} {\bibinfo {author} {\bibfnamefont {A.~M.}\ \bibnamefont
  {Karo}}\ and\ \bibinfo {author} {\bibfnamefont {J.~R.}\ \bibnamefont
  {Hardy}},\ }\href {\doibase 10.1063/1.1669590} {\bibfield  {journal}
  {\bibinfo  {journal} {J. Chem. Phys.}\ }\textbf {\bibinfo {volume} {48}},\
  \bibinfo {pages} {3173} (\bibinfo {year} {1968})}\BibitemShut {NoStop}%
\bibitem [{\citenamefont {Spassky}\ \emph {et~al.}(2015)\citenamefont
  {Spassky}, \citenamefont {Nagirnyi}, \citenamefont {Savon}, \citenamefont
  {Kamenskikh}, \citenamefont {Barinova}, \citenamefont {Kirsanova},
  \citenamefont {Grigorieva}, \citenamefont {Ivannikova}, \citenamefont
  {Shlegel}, \citenamefont {Aleksanyan},\ and\ \citenamefont
  {et~al.}}]{LiMoO_bandgapvsT}%
  \BibitemOpen
  \bibfield  {author} {\bibinfo {author} {\bibfnamefont {D.}~\bibnamefont
  {Spassky}}, \bibinfo {author} {\bibfnamefont {V.}~\bibnamefont {Nagirnyi}},
  \bibinfo {author} {\bibfnamefont {A.}~\bibnamefont {Savon}}, \bibinfo
  {author} {\bibfnamefont {I.}~\bibnamefont {Kamenskikh}}, \bibinfo {author}
  {\bibfnamefont {O.}~\bibnamefont {Barinova}}, \bibinfo {author}
  {\bibfnamefont {S.}~\bibnamefont {Kirsanova}}, \bibinfo {author}
  {\bibfnamefont {V.}~\bibnamefont {Grigorieva}}, \bibinfo {author}
  {\bibfnamefont {N.}~\bibnamefont {Ivannikova}}, \bibinfo {author}
  {\bibfnamefont {V.}~\bibnamefont {Shlegel}}, \bibinfo {author} {\bibfnamefont
  {E.}~\bibnamefont {Aleksanyan}}, \ and\ \bibinfo {author} {\bibnamefont
  {et~al.}},\ }\href {\doibase 10.1016/j.jlumin.2015.05.042} {\bibfield
  {journal} {\bibinfo  {journal} {J. Lumin.}\ }\textbf {\bibinfo {volume}
  {166}},\ \bibinfo {pages} {195–202} (\bibinfo {year} {2015})}\BibitemShut
  {NoStop}%
\bibitem [{\citenamefont {Smith}\ \emph {et~al.}(1996)\citenamefont {Smith},
  \citenamefont {Arnison}, \citenamefont {Homer}, \citenamefont {Lewin},
  \citenamefont {Alner}, \citenamefont {Spooner}, \citenamefont {Quenby},
  \citenamefont {Sumner}, \citenamefont {Bewick}, \citenamefont {Li},
  \citenamefont {Shaul}, \citenamefont {Ali}, \citenamefont {Jones},
  \citenamefont {Smith}, \citenamefont {Davies}, \citenamefont {Lally},
  \citenamefont {{van den Putte}}, \citenamefont {Barton},\ and\ \citenamefont
  {Blake}}]{decayPSD}%
  \BibitemOpen
  \bibfield  {author} {\bibinfo {author} {\bibfnamefont {P.}~\bibnamefont
  {Smith}}, \bibinfo {author} {\bibfnamefont {G.}~\bibnamefont {Arnison}},
  \bibinfo {author} {\bibfnamefont {G.}~\bibnamefont {Homer}}, \bibinfo
  {author} {\bibfnamefont {J.}~\bibnamefont {Lewin}}, \bibinfo {author}
  {\bibfnamefont {G.}~\bibnamefont {Alner}}, \bibinfo {author} {\bibfnamefont
  {N.}~\bibnamefont {Spooner}}, \bibinfo {author} {\bibfnamefont
  {J.}~\bibnamefont {Quenby}}, \bibinfo {author} {\bibfnamefont
  {T.}~\bibnamefont {Sumner}}, \bibinfo {author} {\bibfnamefont
  {A.}~\bibnamefont {Bewick}}, \bibinfo {author} {\bibfnamefont
  {J.}~\bibnamefont {Li}}, \bibinfo {author} {\bibfnamefont {D.}~\bibnamefont
  {Shaul}}, \bibinfo {author} {\bibfnamefont {T.}~\bibnamefont {Ali}}, \bibinfo
  {author} {\bibfnamefont {W.}~\bibnamefont {Jones}}, \bibinfo {author}
  {\bibfnamefont {N.}~\bibnamefont {Smith}}, \bibinfo {author} {\bibfnamefont
  {G.}~\bibnamefont {Davies}}, \bibinfo {author} {\bibfnamefont
  {C.}~\bibnamefont {Lally}}, \bibinfo {author} {\bibfnamefont
  {M.}~\bibnamefont {{van den Putte}}}, \bibinfo {author} {\bibfnamefont
  {J.}~\bibnamefont {Barton}}, \ and\ \bibinfo {author} {\bibfnamefont
  {P.}~\bibnamefont {Blake}},\ }\href {\doibase
  https://doi.org/10.1016/0370-2693(96)00350-4} {\bibfield  {journal} {\bibinfo
   {journal} {Phys. Lett. B}\ }\textbf {\bibinfo {volume} {379}},\ \bibinfo
  {pages} {299 } (\bibinfo {year} {1996})}\BibitemShut {NoStop}%
\bibitem [{\citenamefont {Lu}\ \emph {et~al.}(2015)\citenamefont {Lu},
  \citenamefont {Li}, \citenamefont {Bizarri}, \citenamefont {Yang},
  \citenamefont {Mayhugh}, \citenamefont {Menge},\ and\ \citenamefont
  {Williams}}]{Lu_CsIRmT}%
  \BibitemOpen
  \bibfield  {author} {\bibinfo {author} {\bibfnamefont {X.}~\bibnamefont
  {Lu}}, \bibinfo {author} {\bibfnamefont {Q.}~\bibnamefont {Li}}, \bibinfo
  {author} {\bibfnamefont {G.~A.}\ \bibnamefont {Bizarri}}, \bibinfo {author}
  {\bibfnamefont {K.}~\bibnamefont {Yang}}, \bibinfo {author} {\bibfnamefont
  {M.~R.}\ \bibnamefont {Mayhugh}}, \bibinfo {author} {\bibfnamefont {P.~R.}\
  \bibnamefont {Menge}}, \ and\ \bibinfo {author} {\bibfnamefont {R.~T.}\
  \bibnamefont {Williams}},\ }\href {\doibase 10.1103/PhysRevB.92.115207}
  {\bibfield  {journal} {\bibinfo  {journal} {Phys. Rev. B}\ }\textbf {\bibinfo
  {volume} {92}},\ \bibinfo {pages} {115207} (\bibinfo {year}
  {2015})}\BibitemShut {NoStop}%
\bibitem [{\citenamefont {Salakhutdinov}\ and\ \citenamefont
  {Efanov}(2015)}]{lotofscin}%
  \BibitemOpen
  \bibfield  {author} {\bibinfo {author} {\bibfnamefont {G.}~\bibnamefont
  {Salakhutdinov}}\ and\ \bibinfo {author} {\bibfnamefont {D.}~\bibnamefont
  {Efanov}},\ }\href {https://doi.org/10.1134/S0020441215030100} {\ \textbf
  {\bibinfo {volume} {58}},\ \bibinfo {pages} {345} (\bibinfo {year}
  {2015})}\BibitemShut {NoStop}%
\bibitem [{\citenamefont {Donghia}(2016)}]{mu2_csi}%
  \BibitemOpen
  \bibfield  {author} {\bibinfo {author} {\bibfnamefont {R.}~\bibnamefont
  {Donghia}},\ }\href@noop {} {\bibfield  {journal} {\bibinfo  {journal} {Nuovo
  Cimento C}\ }\textbf {\bibinfo {volume} {39}},\ \bibinfo {pages} {276}
  (\bibinfo {year} {2016})}\BibitemShut {NoStop}%
\bibitem [{\citenamefont {Scholz}\ \emph {et~al.}(2016)\citenamefont {Scholz},
  \citenamefont {Chavarria}, \citenamefont {Collar}, \citenamefont
  {Privitera},\ and\ \citenamefont {Robinson}}]{Scholz_GEconYBe}%
  \BibitemOpen
  \bibfield  {author} {\bibinfo {author} {\bibfnamefont {B.}~\bibnamefont
  {Scholz}}, \bibinfo {author} {\bibfnamefont {A.}~\bibnamefont {Chavarria}},
  \bibinfo {author} {\bibfnamefont {J.}~\bibnamefont {Collar}}, \bibinfo
  {author} {\bibfnamefont {P.}~\bibnamefont {Privitera}}, \ and\ \bibinfo
  {author} {\bibfnamefont {A.}~\bibnamefont {Robinson}},\ }\href
  {http://dx.doi.org/10.1103/PhysRevD.94.122003} {\bibfield  {journal}
  {\bibinfo  {journal} {Phys. Rev. D}\ }\textbf {\bibinfo {volume} {94}}
  (\bibinfo {year} {2016})}\BibitemShut {NoStop}%
\bibitem [{\citenamefont {Collar}(2013{\natexlab{b}})}]{Collar_YBe}%
  \BibitemOpen
  \bibfield  {author} {\bibinfo {author} {\bibfnamefont {J.~I.}\ \bibnamefont
  {Collar}},\ }\href@noop {} {\bibfield  {journal} {\bibinfo  {journal} {Phys.
  Rev. Lett.}\ }\textbf {\bibinfo {volume} {110 21}},\ \bibinfo {pages}
  {211101} (\bibinfo {year} {2013}{\natexlab{b}})}\BibitemShut {NoStop}%
\bibitem [{\citenamefont {Chavarria}\ \emph {et~al.}(2016)\citenamefont
  {Chavarria} \emph {et~al.}}]{alvaro}%
  \BibitemOpen
  \bibfield  {author} {\bibinfo {author} {\bibfnamefont {A.~E.}\ \bibnamefont
  {Chavarria}} \emph {et~al.},\ }\href {\doibase 10.1103/PhysRevD.94.082007}
  {\bibfield  {journal} {\bibinfo  {journal} {Phys. Rev. D}\ }\textbf {\bibinfo
  {volume} {94}},\ \bibinfo {pages} {082007} (\bibinfo {year}
  {2016})}\BibitemShut {NoStop}%
\bibitem [{\citenamefont {Ibe}\ \emph {et~al.}(2017)\citenamefont {Ibe},
  \citenamefont {Nakano}, \citenamefont {Shoji},\ and\ \citenamefont
  {Suzuki}}]{Migdal}%
  \BibitemOpen
  \bibfield  {author} {\bibinfo {author} {\bibfnamefont {M.}~\bibnamefont
  {Ibe}}, \bibinfo {author} {\bibfnamefont {W.}~\bibnamefont {Nakano}},
  \bibinfo {author} {\bibfnamefont {Y.}~\bibnamefont {Shoji}}, \ and\ \bibinfo
  {author} {\bibfnamefont {K.}~\bibnamefont {Suzuki}},\ }\href@noop {}
  {\bibfield  {journal} {\bibinfo  {journal} {JHEP}\ }\textbf {\bibinfo
  {volume} {2018}},\ \bibinfo {pages} {1} (\bibinfo {year} {2017})}\BibitemShut
  {NoStop}%
\bibitem [{\citenamefont {Kahn}\ \emph {et~al.}(2020)\citenamefont {Kahn},
  \citenamefont {Krnjaic},\ and\ \citenamefont {Mandava}}]{yoni}%
  \BibitemOpen
  \bibfield  {author} {\bibinfo {author} {\bibfnamefont {Y.}~\bibnamefont
  {Kahn}}, \bibinfo {author} {\bibfnamefont {G.}~\bibnamefont {Krnjaic}}, \
  and\ \bibinfo {author} {\bibfnamefont {B.}~\bibnamefont {Mandava}},\
  }\href@noop {} {\  (\bibinfo {year} {2020})},\ \Eprint
  {http://arxiv.org/abs/2011.09477} {arXiv:2011.09477 [hep-ph]} \BibitemShut
  {NoStop}%
\bibitem [{\citenamefont {Knapen}\ \emph {et~al.}(2020)\citenamefont {Knapen},
  \citenamefont {Kozaczuk},\ and\ \citenamefont {Lin}}]{tongyan}%
  \BibitemOpen
  \bibfield  {author} {\bibinfo {author} {\bibfnamefont {S.}~\bibnamefont
  {Knapen}}, \bibinfo {author} {\bibfnamefont {J.}~\bibnamefont {Kozaczuk}}, \
  and\ \bibinfo {author} {\bibfnamefont {T.}~\bibnamefont {Lin}},\ }\href@noop
  {} {\  (\bibinfo {year} {2020})},\ \Eprint {http://arxiv.org/abs/2011.09496}
  {arXiv:2011.09496 [hep-ph]} \BibitemShut {NoStop}%
\end{thebibliography}%

\end{document}